\documentclass[reprint, amsmath,amssymb,onecolumn]{revtex4-2}
\usepackage{setspace}
\usepackage{graphicx}
\usepackage{dcolumn}
\usepackage{bm}
\usepackage{lineno,hyperref}
\usepackage{xcolor}
\usepackage{physics}
\usepackage[version=4]{mhchem}
\modulolinenumbers[5]
\hypersetup{colorlinks = true,linkcolor = blue,anchorcolor =red,citecolor = blue,filecolor = red,urlcolor = black,
            pdfauthor=author}

\providecommand\caj{C^a_j}
\providecommand\cwj{C^w_j}
\providecommand\Daj{D^a_j}
\providecommand\Dwj{D^w_j}
\providecommand\nbj{n^b_j}
\providecommand\kaj{S^a_j}
\providecommand\kwj{S^w_j}
\providecommand\pbj{P^b_j}

\providecommand\peq{P_\mathit{eq}}
\providecommand\Veq{V_\mathit{eq}}
\providecommand\oeq{^{0,\mathit{eq}}}
\providecommand\Raf{\mathit{Ra_f}}
\providecommand\Ra{\mathit{Ra_w}}
\providecommand\Rac{\mathit{Ra_c}}
\providecommand\Raa{\mathit{Ra_a}}
\providecommand\Sh{\mathit{Sh_w}}
\providecommand\Sc{\mathit{Sc}}
\newcommand\Pra{\mathit{Pr}}
\providecommand\cw{_\mathrm{CO_2}^w}
\providecommand\co{_\mathrm{CO_2}^0}
\providecommand\cb{_\mathrm{CO_2}^b}
\providecommand\ca{_\mathrm{CO_2}^a}
\providecommand\cm{_\mathrm{CO_2}^{a/w}}

\providecommand\jwb{^{wb}_j}
\providecommand\jba{^{ba}_j}
\providecommand\cwb{^{wb}_\mathrm{CO_2}}
\providecommand\cba{^{ba}_\mathrm{CO_2}}
\newcommand\ee{\mathrm{e}^}
\newcommand\erfc{\mathrm{erfc}}

\begin{document}



\title{Slug bubble growth and dissolution by solute exchange}

\author{Dani\"{e}l P. Faasen}
   \email{d.p.faasen@utwente.nl}
\author{Devaraj van der Meer}%
\email{d.vandermeer@utwente.nl}
\author{Detlef Lohse}%
\email{d.lohse@utwente.nl}
\author{Pablo Pe\~{n}as}%
\email{p.penaslopez@utwente.nl}
\affiliation{%
Physics of Fluids Group, Faculty of Science and Technology, University of Twente, P.O. Box 217, 7500 AE Enschede, The Netherlands\\}%
\date{\today}

\begin{abstract}

In many environmental and industrial applications, the mass transfer of gases in liquid solvents is a fundamental process during the generation of bubbles for specific purposes or, vice versa, the removal of entrapped bubbles. We address the growth dynamics of a trapped slug bubble in a vertical glass cylinder under a water barrier. In the studied process, the ambient air atmosphere is replaced by a CO$_2$ atmosphere at the same or higher pressure. 
The asymmetric exchange of the gaseous solutes between the CO$_2$-rich water barrier and the air-rich bubble always results in net bubble growth.  We refer to this process  as solute exchange. 
The dominant transport of CO$_2$ across the water barrier is driven by a combination of diffusion and convective dissolution. 
The experimental results are compared to and explained with a simple numerical model, with which the underlying mass transport processes are quantified. Analytical solutions that accurately predict  the bubble growth dynamics are subsequently derived. The effect of convective dissolution across the water layer is treated as a reduction of the effective diffusion length, in accordance with the mass transfer scaling observed in laminar or natural convection.
 Finally, the binary water--bubble  system is extended to a ternary water--bubble--alkane system.  It is found that the alkane  (n-hexadecane) layer bestows a buffering (hindering) effect on bubble growth and dissolution. The resulting growth dynamics and underlying fluxes are characterised theoretically.

\end{abstract}

\maketitle

\section{Introduction}

The dissolution of carbon dioxide gas and its subsequent transport through liquid layers plays an important role in many  industrial and biological applications. Examples include enhanced oil recovery  \cite{Jiang2019}, CO$_2$ sequestration operations in porous media \cite{Huppert2014}, and gas exchange in the respiratory system  \cite{Grotberg2011}. In microfluidics, segmented flows of alternating
 CO$_2$ gas and liquid plugs are frequently employed to study solvent absorption properties and chemical reaction kinetics \cite{Cao2021, Abolhasani2014, Cubaud2012}. In fact, the controlled dissolution of CO$_2$ bubbles constitutes a crucial step in the microfluidic generation of  encapsulated microbubbles for ultrasound-imaging in medical diagnostics and targeted drug delivery \cite{Lu2016, Stride2008}.
  
Bubble formation, however, can have detrimental effects on microfluidic devices such as blocking the liquid flow \cite{Pereiro2019, Volk2015}. Bubbles covering the electrode surface of electrochemical devices hinder gas-evolution reactions, thereby reducing the efficiency of the cell \cite{Espinosa2018, Prakash2008, Meng2007, Litterst2006, Pande2020}. Similarly, bubble formation during inkjet printing can completely disrupt the printing process \cite{Lohse2022}. Therefore, thorough understanding of the solubility and transport of  dissolved gas is clearly of benefit, not only upon encountering these complications, but also in an overwhelming number of applications concerning gas or vapor bubbles  \cite{Lohse2018}.

Multiple solutions for unwanted bubble removal exist, ranging from passive removal such as bubble traps, to active removal by increasing the driving pressure in order to increase the gas solubility and induce dissolution \cite{Pereiro2019}. However, when performing the latter, one should take note of the gas composition inside the trapped bubble and surrounding liquid, as pressurising with a different gas species may very well cause the bubble to grow instead of dissolve.

This is precisely the focus of our work. We investigate the mass transfer experienced by a single slug air bubble trapped in a vertical glass cylinder beneath a thin layer of water which separates the bubble from the ambient gas, as shown in Fig.~\ref{FIG:1}. We replace the ambient air atmosphere with a CO$_2$ atmosphere at the same or higher pressure, and study the subsequent growth dynamics of the bubble. Its growth is driven by the asymmetric exchange of gaseous solutes in the water layer (CO$_2$) with those in the trapped bubble (air). We therefore refer to this driving mechanism as solute exchange, where by replacing the gaseous solutes in a liquid solvent one is able to promote the growth (or dissolution) of the bubbles found therein. Solute exchange is fundamentally different from the solvent exchange mechanism used to generate microbubbles by replacing the liquid solvents \cite{Peng2016, Lu2015}. 

In a nutshell, our aim is to investigate the growth dynamics of a trapped slug bubble due to a solute exchange process induced by changes in the outer ambient gas composition. 
It will be seen that the bubble growth rate is in fact supradiffusive due to the onset of  CO$_2$ dissolution-driven convection across the vertical water layer above the bubble. 
Convective dissolution or growth is indeed encountered  in many relevant applications, such as CO$_2$ capture and storage in saline aquifers \cite{Huppert2014}, buoyancy-generating chemical reactions \cite{Rogers2005, Loodts2018}, droplet dissolution \cite{Dietrich2016, Chong2020} or bubble growth \cite{Soto2019}. In essence, our system embodies a mass-transport variant of the classical problem of Rayleigh--B{\'e}nard convection \cite{Ahlers2009} in a cylindrical cell. With this in mind, we devote significant effort to the quantification, both experimentally and theoretically, of the effect of convective dissolution on the bubble growth dynamics. Finally, the binary water--bubble system is extended to a ternary water--bubble--n-hexadecane system. Bubble growth dynamics will be seen to be hindered by the presence of the buffering alkane layer. Ternary configurations are of considerable relevance in gas--liquid--liquid three-phase microreactors \cite{Yao2017}. The inclusion of oxygenated oil layers has also been shown to enhance bacterial growth in bioreactors \cite{Sklodowska2017}, owing to the well-known fact that oils have higher gas solubilities than water \cite{Sander2015, Smith2007, Makranczy1976}.

This paper is structured as follows. The experimental set-up and procedure are described in section \ref{sec:exp}. Section \ref{sec:solutexchange} offers a  qualitatively description of the solute exchange mechanism. Section \ref{sec:wb} presents quantitative results for the aforementioned binary water--bubble system. We propose a simple theoretical framework which adequately describes the bubble growth dynamics driven by (convection-enhanced) solute exchange. Experiments are compared to a numerical model and analytical solutions. The proposed dependence of the Sherwood number on the Rayleigh number across the water layer is compared to all binary and ternary experiments in section \ref{sec:scaling}.
In section \ref{sec:wba}  the ternary water--bubble--alkane system is investigated in greater detail, and the  buffering effect of  the n-hexadecane layer is characterised. 
The paper ends with a summary of the main findings and an outlook in section \ref{sec:conc}. 

\section{Experimental set-up and procedure for solute exchange}
\label{sec:exp}

\begin{figure*}
	\centering
	\includegraphics[width=0.8\textwidth]{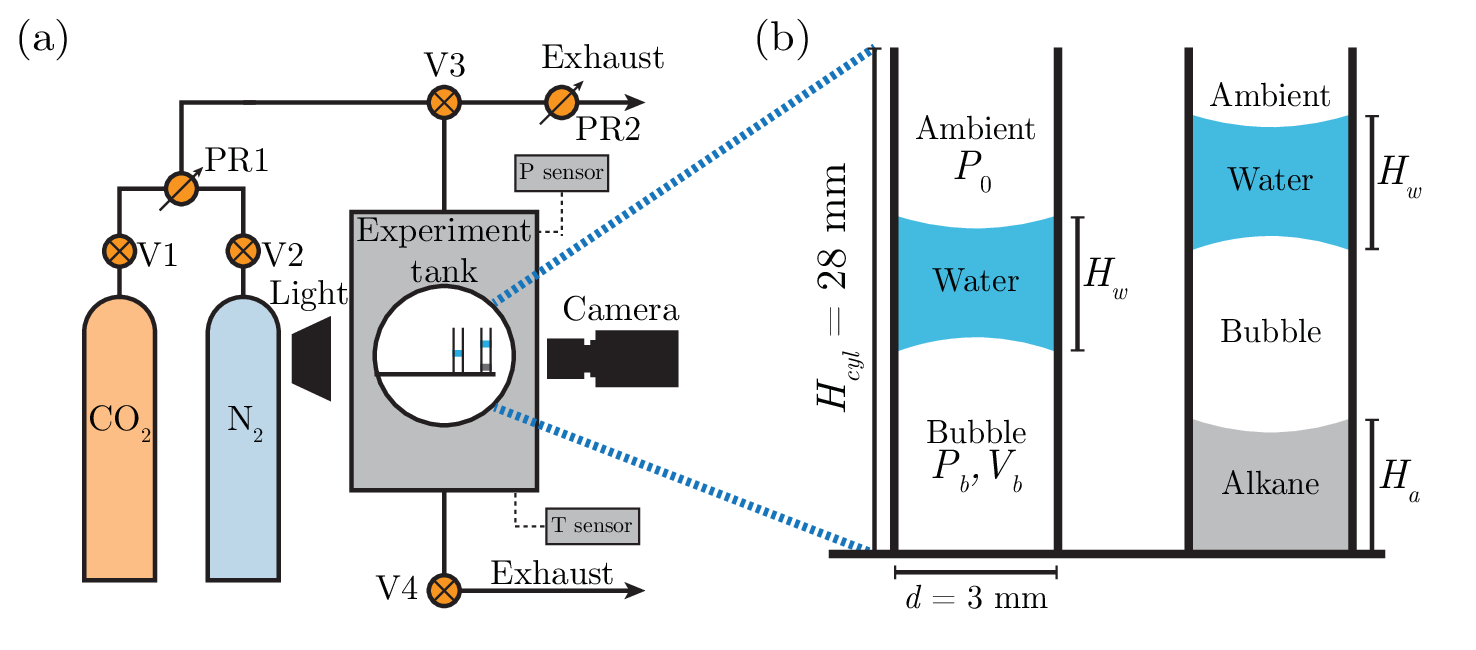}
	\caption{(a) Schematic overview of the experimental set-up. (b) Sketch of the cylinders containing the water--bubble and water--bubble-alkane systems. These are placed inside the chamber, which is subsequently pressurised with CO$_2$ gas.}
	\label{FIG:1}
\end{figure*}

A schematic overview of our experimental set-up is shown in Fig.~\ref{FIG:1}(a). The experiments are conducted inside a sealed chamber which can be pressurised with either CO$_2$ or N$_2$ gas. The inlet pressure is adjusted using pressure regulator PR1, whereas a second pressure regulator, PR2, grants additional control of the ambient pressure $P_0$ inside the tank, where $1 \leq P_0 \leq 4$ bar. A temperature and a pressure sensor monitor the temperature $T_0$ and pressure $P_0$ in time. Extensive details on the chamber and pressure control system can be found elsewhere \cite{Enriquez2013}.

Two borosilicate glass (Duran) cylinders (28 mm in length, inner diameter $d = 3.0$ mm, outer diameter of 5.0 mm) are attached on one end to a silicon wafer plate using Loctite 4305 (Farnell), while the remaining end is left open. Before use, the cylinders are rinsed with ethanol (Boom, technical grade) followed by Milli-Q water (resistivity = 18.2 M$\Omega$ $\cdot$ cm) and finally dried in a nitrogen stream. 

The cylinders host either a water--bubble or a  water--bubble--alkane configuration, as depicted in  Fig.~\ref{FIG:1}(b). 
 The water--bubble system is prepared by injecting  a slug bubble using an air-filled syringe between the plate wall and a previously deposited Milli-Q water layer of height $H_w = 2.5$--6.0 mm. The bubble is held in place due to a stable balance between the weight of the water barrier and a combination of the surface tension of the water-bubble interface and to a lesser extent the differences in gas pressures. This force balance persists during the experiments, given that the surface tension of CO$_2$ on water is almost the same as the surface tension of air on water under our experimental conditions \cite{Chun1995}. For the water--bubble--alkane configuration, the bubble is similarly injected between a n-hexadecane  layer (VWR, 99\% purity) of height of $H_a \approx 3.0$ mm and the water layer, whose $H_w$ was adequately chosen from within the same range of lengths  as in the binary system for fair comparison. The selection of bubble volumes will be discussed shortly.

\begin{figure*}
	\centering
	\includegraphics[width=0.7\textwidth]{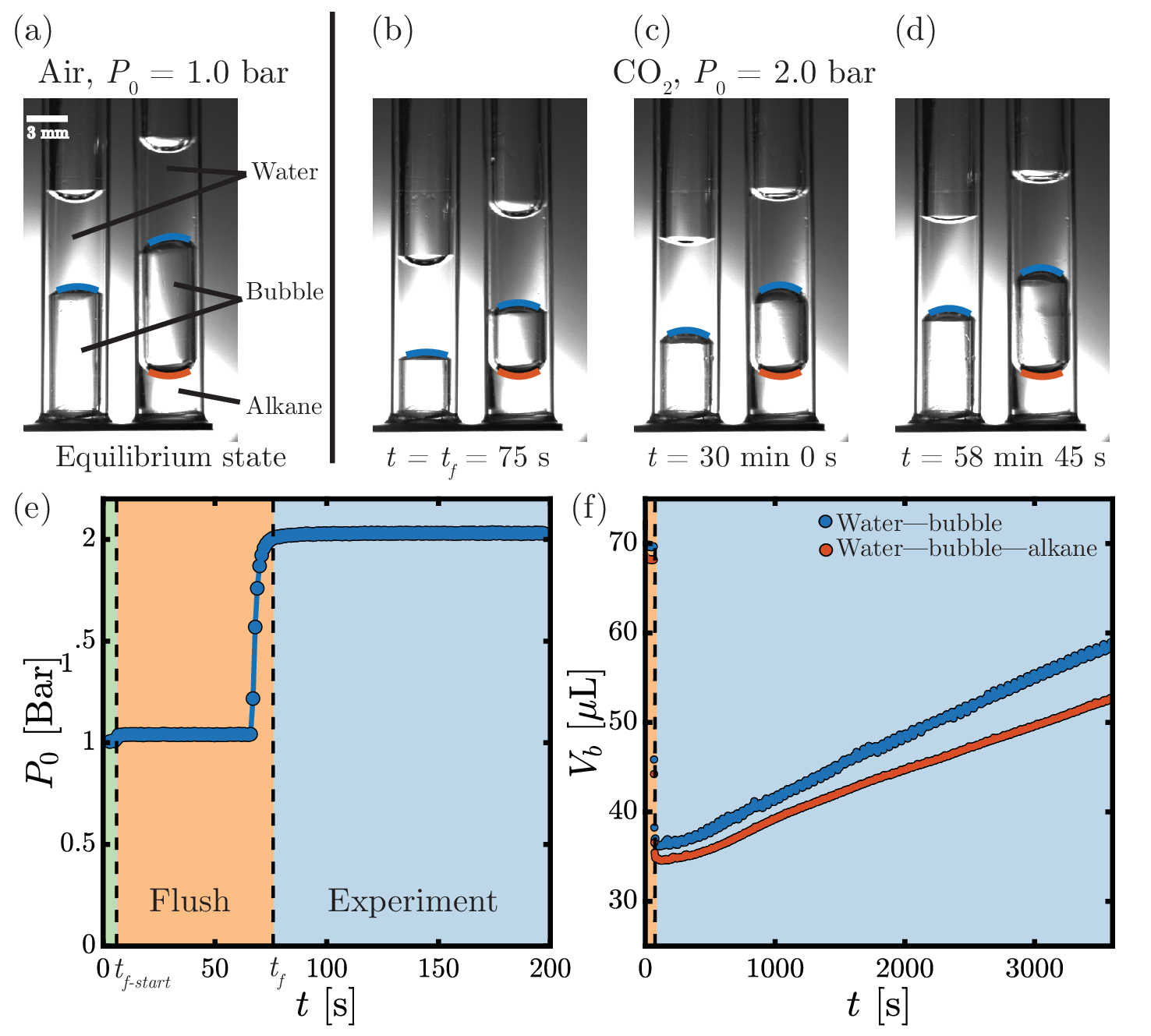}
	\caption{A typical experiment involving a binary and ternary system. Experimental snapshots at (a) $t = 0$, equilibrium state,  (b) $t$ = $t_f$, immediately after CO$_2$-flushing and pressurisation, (c) halfway through the experiment and (d) at the end of the experiment. The blue line indicates the bubble--water interface detected by the contour detection algorithm. Similarly, in the right cylinder with the alkane, the red line indicates the detected  bubble--alkane interface. (e) Time evolution of the pressure in the tank during the first 200 seconds of the experiment. Before $t_{f-start}$ (green stage) the tank is filled with ambient air. Between $t_{f-start}$ and $t_f$ (orange stage) the tank is flushed with CO$_2$ and the bubbles are compressed. 
(f) Evolution in time of the bubble volume  for the two systems.}
	\label{FIG:2}
\end{figure*}

The experimental procedure is best illustrated with reference to a typical experiment shown in  Fig.~\ref{FIG:2}.
The prepared cylinders are placed inside the experimental tank. The inlet pressure is set to 1.0 bar with valve V3 remaining closed. At time $t$ = 0, the camera (Photron FASTCAM Nova s12, 2 fps) and the pressure/temperature sensors (2 aquisitions per second) start recording. At this point, the water--bubble and water--bubble--alkane systems will be in the equilibrium state [Fig.~\ref{FIG:2}(a)]. The volume of any particular bubble, $V_b(t)$, is initially at equilibrium: $V_b(0) \equiv \Veq$.   
At time $t_{f-start}$ = 5 s, valves V3 and V4 [see Fig.~\ref{FIG:1}(a)] are opened and the system is flushed with CO$_2$ gas in order to fully replace the ambient air in the experimental chamber. This corresponds to the ``flush'' stage in Fig.~\ref{FIG:2}(e), which plots the pressure inside the experimental chamber, $P_0(t)$, during the initial minutes of that particular experiment. After flushing for 60 seconds, valve V4 is closed and $P_0$ is set to the  pressure level of the ``experiment'' stage, e.g.,  $P_0$ =  2.0 bar for the experiment in Fig.~\ref{FIG:2}. 
The time $t_f$ [typically 75 s, see Fig.~\ref{FIG:2}(e)] importantly refers to the time immediately after the flushing and compression stages have been completed and marks the start of the ``experiment'' stage. The bubble volume at this point [cf. Fig.~\ref{FIG:2}(b)] is coherently denoted by $V_f \equiv V_b(t_f)$. Thereafter, despite pressurization, both bubbles experience continuous growth -- counterintuitive perhaps -- as a consequence of  solute exchange. The growth is showcased by snapshots in Fig.~\ref{FIG:2}(b--d) and its dynamics is quantified in the $V_b(t)$ plot in Fig.~\ref{FIG:2}(f).
At the end of the experiment (one hour in this case), valves V1--V3 are closed and V4 is opened in order to depressurise the system. The bubble volumes are extracted from the images  using an in-house developed Matlab script.

Several experiments similar to the one just described were performed for binary and ternary systems, or a combination thereof, comprising  either a single-system (one-cylinder) experiment  or a  two-system (two-cylinder) experiment like the one shown in Fig.~\ref{FIG:2}. The pressure level $P_0(t>t_f)$ ranged from 1 bar (no compression after flushing) up to 4 bar,  with  running times of typically one or two hours. The initial bubble volumes considered correspond to (post-pressurization) volumes of 25 $\mu$L $<V_f <45$ $\mu$L (3.5  to 6.4 mm in height). The ambient temperature remained fairly constant at approximately $T_0 = 295$ K for every experiment.

It should be stressed that when pressurising the chamber above the atmospheric pressure [$P_0(t>t_f)>1.0$ bar], the bubble volume compresses from $\Veq$ to $V_f$ accordingly [cf. Fig.~\ref{FIG:2}(a, b)], as established by Boyle's law.  Moreover, the movement of the contact lines during compression also alters the shape of the bubble menisci, which may even take an additional few seconds to reform and settle. In some cases, we were unable to capture  $\Veq$ with sufficient accuracy due to optical limitations. For these reasons, the reference bubble volume is taken to be $V_f$ instead of $\Veq$.
This decision is justified provided that any changes in the molar gas contents in the bubble (initially air) remain negligible during $t<t_f$, i.e., under the assumption that there is insignificant mass transfer to or from the bubble up to that point. 
The experiments conducted at $P_0(t>t_f)$ = 1.0 bar (devoid of pressurisation) fully support the above assumption: $V_f$ was seen to effectively remain unchanged from $\Veq$.

\section{Solute exchange mechanism}
\label{sec:solutexchange}
The replacement of the air atmosphere inside the experimental chamber by CO$_2$ gas causes the latter to dissolve into the water layer, where a CO$_2$-concentration boundary layer naturally develops. Hence, CO$_2$ is transported  down towards the water--bubble interface, where it exsolves into the bubble. The steady-state diffusive influx into the bubble (in the absence of convection) can be estimated noting that
the concentration of dissolved CO$_2$ at the water--bubble interface is initially close to zero, whereas at the water--ambient interface it is $S\cw P_0$, where $S\cw$ denotes the Henry coefficient of  CO$_2$ in water. The flux estimate is therefore $S\cw P_0/H_w$, where $H_w$ is the height of the water column, see Fig.~\ref{FIG:1}.

At the same time, an air concentration gradient forms across the water layer, sustaining a diffusive flux in the opposite direction, i.e., from the (air-rich) bubble to the (air-depleted) CO$_2$ atmosphere. The corresponding diffusive outflux can be estimated in the same manner as $S_\mathit{air}^w P_0/H_w$, where  $S_\mathit{air}^w$ denotes the molar-averaged Henry coefficient of the air mixture  in water. 
The high solubility of CO$_2$ in water (cf.  the solubilities listed in  \autoref{tab:constants}) implies that the CO$_2$  influx is stronger than the air outflux by a factor of $S\cw/S_\mathit{air} ^w\approx 42 $. This clear disparity in the magnitude between both solute exchange fluxes results in net bubble growth as seen in Fig.~\ref{FIG:2}. 

The time required for diffusion to develop a linear CO$_2$-concentration profile across the full height of the water layer can be estimated as $\sim$$0.4 H_w^2/D\cw$ from simple theoretical considerations [see appendix \ref{sec:appA},  Fig.~\ref{FIG:A1}(a)], where $D\cw$ denotes the diffusivity of CO$_2$ in  water.  For  the experiment portrayed in Fig.~\ref{FIG:2} where $H_w = 3$~mm, the bubbles would require an initial transient period of $\sim$2,000~s before the maximum quasi-steady growth rate is attained. However,  Fig.~\ref{FIG:2}(f) clearly shows that the transient period is substantially smaller, of approximately 300~s. 
Thus, without further calculation, it is evident that convective dissolution must have a dominant influence on the bubble growth. Its effect will be unraveled next in Sec. \ref{sec:wb}.

Finally, Fig.~\ref{FIG:2}(f) also reveals that the presence of the n-hexadecane layer slows down bubble growth.  Essentially, part of the CO$_2$ accumulating into the bubble is redissolving into the hexadecane layer, given that  it is initially depleted of CO$_2$. The stabilising or buffering effect of the alkane layer on the bubble growth (and dissolution) dynamics will be discussed in detail in Sec. \ref{sec:wba}.

 \begin{table*}[!htp]
\caption{Henry's constants and diffusion coefficients of the gases in water (superscript $w$) and n-hexadecane (superscript $a$) at 293.0 K and 1.0 bar. The molar-averaged $S_\mathit{air}^w$ can be estimated as $9.02\times 10^{-6}$mol/m$^3$ Pa.}
\label{tab:constants}
\begin{ruledtabular}
\begin{tabular}{lcccc}
      $j$ & $S^{w}_{j}$ [mol/m$^3$ Pa] &$S^{a}_{j}$ [mol/m$^3$ Pa] & $D^w_{j}$ [m$^2$/s] & $D^{a}_{j}$ [m$^2$/s] \\ \hline
CO$_2$ & 3.79 $\times 10^{-4}$ \cite{Sander2015} & 4.06 $\times 10^{-4}$ \cite{Smith2007} & 1.76 $\times 10^{-9}$ \cite{Tamimi1994} & 2.20 $\times 10^{-9\hphantom{aa}}$ \cite{Matthews1987}             \\

O$_2$  & 1.32 $\times 10^{-5}$ \cite{Sander2015} & 6.45 $\times 10^{-5}$ \cite{Makranczy1976} & 2.03 $\times 10^{-9}$ \cite{Zandi1970} & 2.49 $\times 10^{-9\hphantom{a}}$\footnotemark[1]  \cite{Ju1989}             \\

N$_2$  & 6.89 $\times 10^{-6}$ \cite{Sander2015} & 4.61 $\times 10^{-5}$ \cite{Makranczy1976}  & 2.00 $\times 10^{-9}$  \cite{Himmelblau1964}   & 2.49 $\times 10^{-9\hphantom{a}}$\footnotemark[2]\hphantom{ [35]}         
\end{tabular}
\end{ruledtabular}
\footnotetext[1]{Only available at 294.0 K.}
\footnotetext[2]{Approximated value based on the diffusion coefficient of oxygen in n-hexadecane.}
\end{table*}

\section{Solute exchange in the water--bubble system}
\label{sec:wb}

In this section, we focus solely on the experiments concerning the binary (bubble--water) configuration. We present a theoretical framework for diffusion-driven solute exchange with the additional effect of dissolution-driven convection. The insight gained from a simple numerical model allows us to derive analytical solutions that accurately describe the dynamics of  bubble growth.

\subsection{Effect of convective dissolution}

\begin{figure*}
	\centering
	\includegraphics[width=1\textwidth]{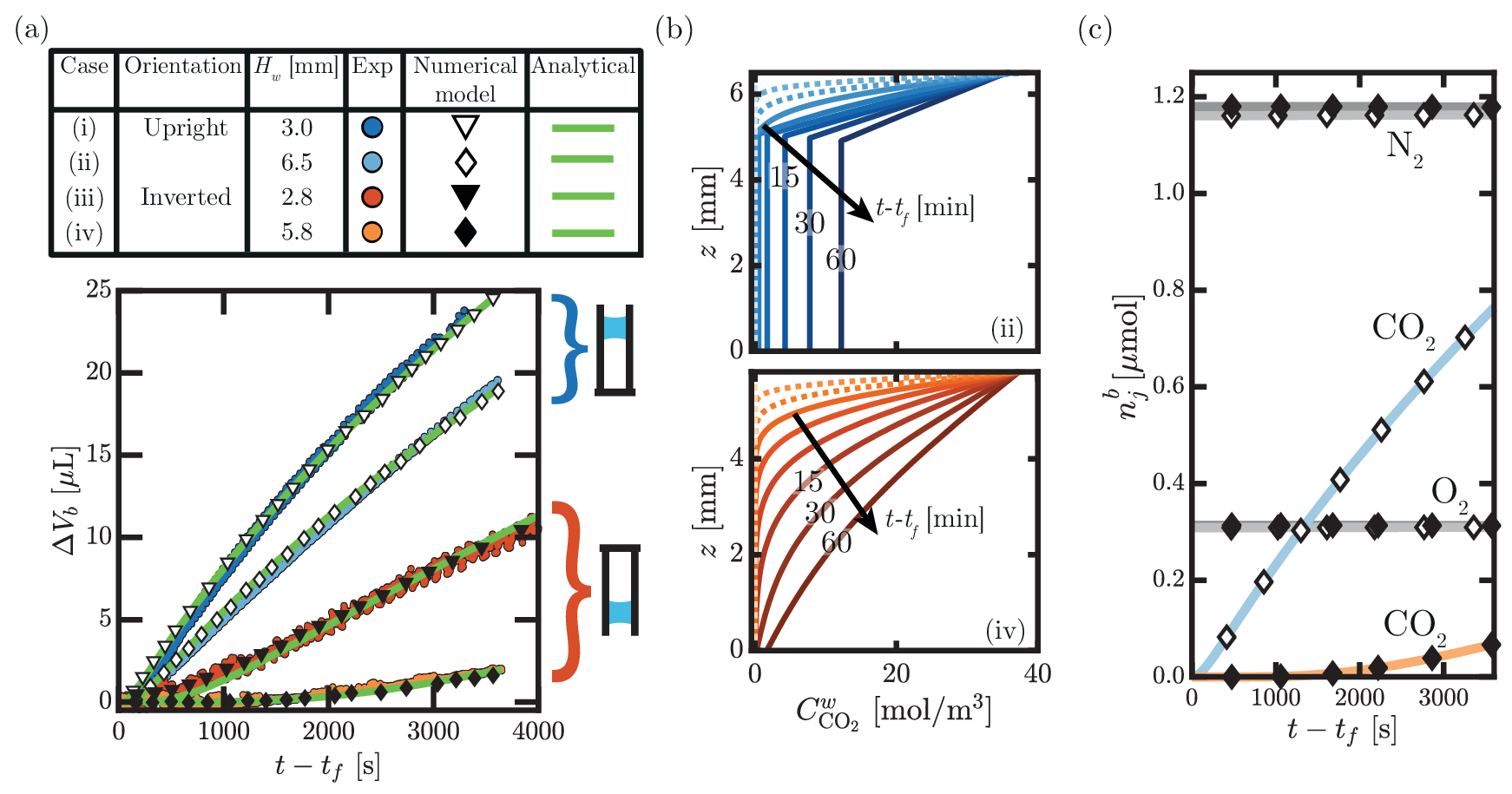}
	\caption{Experimental and theoretical results for water--bubble systems in an upright configuration (cases i, ii) and inverted configuration (cases iii, iv) for two distinct water layer heights; the experiment pressure level is $P_0(t>t_f) = 1$ bar and bubble volumes are initially $V_f = 35$ $\mu$L.
	(a) Bubble growth dynamics ($\Delta V_b \equiv V_b(t)-V_f$), according to experiment (circles), numerical model (black/white markers) and analytical solutions (green lines); the latter are given by (\ref{eq:diffsol})  for (iii, iv) and  (\ref{eq:consol}) for (i, ii).
	The onset of convective dissolution for upright cases (i, ii) is evident.
	(b) Time evolution of the CO$_2$ concentration profiles in the water layer according to the model  for convective dissolution case (ii) and pure diffusion case (iv). The uniform profile outside the diffusion boundary layer  in case (ii)  illustrates the model assumption of instant convective transport.
 (c) Time evolution of the gas composition in the bubble according to the model for upright case (ii) [white markers and lighter curves], and inverted case (iv) [black markers and darker curves].}
	\label{FIG:3}
\end{figure*}

We begin analysing a series of single-cylinder, water--bubble experiments concerning a bubble of  equilibrium volume $V_{eq}$ = 35 $\mu$L capped by a water layer with a thickness of either $H_w \approx 3$ mm or $H_w \approx 6$ mm. The prepared cylinders were placed in the experimental chamber, positioned either upright or inverted (rotated 180$^\circ$), as illustrated in Fig.~\ref{FIG:3}(a). Subsequently, the air inside the chamber was replaced by a CO$_2$ atmosphere at $P_0(t>t_f) = 1 $ bar. Therefore, the bubble volume at the start of the ``experiment'' stage,  $V_f \equiv V_b(t_f)$, is equal to $V_{eq}$.  

The increase in bubble volume with time, $\Delta V_b(t)\equiv V_b(t)-V_f$, is shown for both the upright cases (i, ii) and inverted cases (iii, iv)  in Fig.~\ref{FIG:3}(a). 
The bubbles in the upright configuration are seen to grow much faster than when inverted under otherwise the same conditions. As a consequence, this difference should be attributed to the effect of convective dissolution.
In the upright cases, the dense CO$_2$-rich boundary layer that forms at the top of the water phase becomes gravitationally unstable. This gives rise to dissolution-driven convection in the form of  CO$_2$-rich viscous fingers or plumes that propagate downwards at  relatively high speeds \cite{Backhaus2011,ARENDT2004,Huppert2014, Slim2013, Farajzadeh2007_numerical}. In contrast, the CO$_2$-water mixture is stably stratified when the cylinder is placed inverted and therefore CO$_2$ can only be transported up the water layer by diffusion.

In the absence of convection,  the diffusive transport of the solute exchange (and consequently the bubble growth dynamics) is governed by the one-dimensional diffusion equation for the molar concentration,
\begin{equation}
    \pdv{C^{w}_{j}}{t} = \Dwj \pdv{^2\cwj}{z^2},
    \label{eq:1D-diffusion}
\end{equation} 
across the water layer ($0<z <H_w$) for three gas species: $j = \ce{CO2}$, $j = \ce{O2}$, and $j = \ce{N2}$, where the latter two represent the air mixture. 
In doing so, we neglect the three-dimensional curvature of the bubble meniscus, whose height is indeed small in comparison with $H_w$ [cf. Fig. \ref{FIG:2}(a--d)]. 
The diffusivity of gas $j$ in  water is denoted by $\Dwj$. The diffusive fluxes across the water--bubble interface  on $z = 0$ are therefore $J\jwb = -\Dwj\partial_z \cwj(0, \ t)$.
The ambient pressure $P_0(t)$ and the partial pressures in the ambient are prescribed at all times, and these can be interpolated and inferred from the experimental pressure data. The pressure inside the bubble $P_b(t)$ is assumed equal to $P_0(t)$ at all times, given that the capillary overpressure in the bubble is negligible: $4\sigma/d\ll P_0$, with $\sigma$ as the surface tension. The partial gas pressures in the ambient and the bubble set the interfacial concentrations (boundary conditions) of Eq.~(\ref{eq:1D-diffusion}) as established by Henry's law (see appendix \ref{sec:appB}).
The bubble contents are treated as a mixture of ideal gases, whence $P_0 V_0=n_b RT_0$, where $R$ is the universal gas constant and   $n_b = n^b_\mathrm{CO_2}+n^b_\mathrm{N_2}+n^b_\mathrm{O_2}$ the number of moles of gas in the bubble (neglecting other gases which are present in small amounts ($< 1\%$)). Finally mass conservation,  $J\jwb\pi d^2 /4= \dot n_j^b$, yields the volume growth rate:
 \begin{equation}
    \dot V_b =  \frac{\pi d^2}{4} \frac{RT_0}{P_0} \sum_j J\jwb.
    \label{eq:Vbdot}
\end{equation} 
The above system (\ref{eq:1D-diffusion})--(\ref{eq:Vbdot}) is solved and integrated numerically by means of a finite-differences scheme. A complete description of the model equations, initial conditions and numerical implementation can be found in appendix \ref{sec:appA}.  The resulting time evolution $V_b(t)$ is  plotted by the black markers in  Fig.~\ref{FIG:3}(a) for cases (iii) and (iv). Good agreement is found between these and experiments.

It should be pointed out that we neglect the influence of water vapor (evaporation) on the bubble growth, despite the likely fact that the bubble is saturated with water vapor. This is justified since the water vapour pressure is small, $P_w^\mathit{sat}/P_0 <2$ \% and its inclusion in the theoretical framework would add little insight at the cost of added complexity.
Water evaporation, unlike solute exchange,  is not a driving force for bubble growth. Instead, the role of evaporation is merely to ensure that the bubble remains saturated with water vapour at all times.  It can be shown that evaporation fluxes amplify $\dot V_b(t)$ by a factor $(1-P_w^\mathit{sat}/P_0 )^{-1}$; the resulting difference still remains within the experimental uncertainty of $V_b$.

Next, we incorporate the effect of convective dissolution into the solute exchange diffusive framework established so far. Convective transport of CO$_2$ across the water layer is characterised by the instantaneous Rayleigh and Sherwood numbers,
\begin{equation}
    \label{eq:RaSh}
    \Ra(t) \equiv \frac{\lambda\cw\Delta C\cw(t)gH_w^3}{\nu_w D\cw }, \quad
    \Sh(t) \equiv \frac{J\cwb(t) H_w}{D\cw  \Delta C\cw (t)},
\end{equation}
where $g$ is the acceleration due to gravity, $\nu_w$ the kinematic viscosity of water and $\lambda\cw  \approx 8.2$ cm$^3$/mol the solutal expansion coefficient of CO$_2$ in water at $T_0 = 295$ K \cite{Loodts2014,Garcia2001}. The instantaneous concentration difference across the water layer is established by the partial pressure difference of CO$_2$ gas that exists between the ambient and the bubble: $\Delta C\cw(t)  = S\cw  [P\co(t)-P\cb(t)]$.

In view of the modest aspect ratios of the water column  considered in our experiments ($0.45< d/H_w <1.3$),  we have adopted $H_w$ as the relevant length scale, in agreement with the classical definition of the Rayleigh number in the field of Rayleigh--B\'enard convection \cite{Grossmann2000, Bejan2013}. Here it is worth noting that in contrast to the above choice, convective dissolution studies often employ porous Rayleigh and Sherwood numbers based on the permeability of the porous medium \cite{Huppert2014, Backhaus2011, Kneafsey2011, Slim2013}. There, the permeability scales as $d^2$, namely, the square of the highly restrictive pore size or the Hele-Shaw cell thickness.

The characteristic Rayleigh number can be readily estimated from Eq.~(\ref{eq:RaSh}) inserting  $H_w = 3.0$ mm, $P\co-P\cb = 1.0$ bar in addition to the diffusivities and solubilities tabulated in \autoref{tab:constants}. We obtain $\Ra \approx 3 \times 10^4$, which is well above the critical value, $\Rac = 1708$,  for the onset of convection in an infinitely wide enclosure ($d/H_w\rightarrow \infty$)
\cite{Bejan2013, Hebert2010}.
The Schmidt number is large: $\Sc = \nu_w/D\cw = 568$. 
Therefore, our moderate-$\Ra$, high-$\Sc$ experiments should lie in the viscosity-dominated regime (also large-Prandtl or laminar regime \cite{Bejan2013}, or regime $I_u$ in the Grossmann--Lohse theory \cite{Grossmann2000}), where buoyancy forces are indeed balanced by viscous forces.
We therefore expect a $\Sh \sim \Ra^{1/4}$ dependence \cite{Grossmann2000, Bejan2013}, consistent with measurements of laminar Rayleigh--B\'enard convection \cite{Malkus1954b}, and natural convection adjacent to horizontal surfaces \cite{Goldstein1973, Lloyd1974}, dissolving drops \cite{Dietrich2016, Chong2020} or growing bubbles \cite{Soto2019}. 
This laminar regime should not be mistaken with that of viscous convection of an infinite-Prandtl or Schmidt fluid in the limit of large Rayleigh numbers. For such a case, the scaling $\Sh\sim \Ra^{1/5}$ is expected between (no-slip) solid surfaces \cite{Roberts1979, Vynnycky2013} (cf. regime $I_\infty^>$ of the G--L theory \cite{Grossmann2001}) and $\Sh\sim\Ra^{1/3}$ between (zero shear stress) free surfaces \cite{Roberts1979, Jimenez1987, Vynnycky2013}. However, the robustness of the 1/4 exponent present in all the aforementioned cases of laminar convection hints that neither the no-shear boundary condition nor the cell geometry impact the scaling significantly. In our case, this is further justified by noting that the aspect ratio remains always close to unity, and the fact that bubble surfaces are prone to contamination by impurities naturally present in the ambient and can consequently behave as no-slip \cite{Maali2017}. Nevertheless, the prescribed scaling $\Sh\sim \Ra^{1/4}$ will be verified experimentally later in section \ref{sec:scaling}. 

Provided that variations in $\Ra(t)$ are sufficiently slow, we assume that the dependance of the quasi-steady Sherwood number on the instantaneous Rayleigh number is of the form
  \begin{equation}
\Sh(t) = \left[ 1+ k \frac{\Ra(t)}{\Rac}\right]^{1/4},
\label{eq:Shmodel}
\end{equation}
  where
$\Rac = 1,708$ and $k = 2.75 \pm 1.25 $ is a dimensionless fitting coefficient that can be systematically obtained from the experimental measurements, as will be shown later in section \ref{sec:scaling}. The fitting coefficient $k$ is expected to depend on $\Sc$ and the aspect ratio $d/H_w$ of the system but otherwise it is ideally a constant. Its variability across the experiments arises from several factors such as the different aspect ratios employed,  uneven shapes of the bubble meniscii, or the heterogeneous wetting state of the cylinder walls. In short,  part of the variability is bound to the experimental uncertainty involving $V_b(t)$.
Furthermore, the requirement that (\ref{eq:Shmodel}) should well capture the transition from pure diffusive transport ($\Sh = 1$  for $\Ra < \Rac$) to convection-dominated transport ($\Sh \sim \Ra^{1/4}$ for $ \Ra \gg \Rac $) is not at all important here, since all of our experiments correspond exclusively to the latter. To model details of the transition, an additional fitting parameter could be employed cf. Ref. \cite{Dietrich2016}.
 
\begin{figure}
	\centering
	\includegraphics[width=0.8\textwidth]{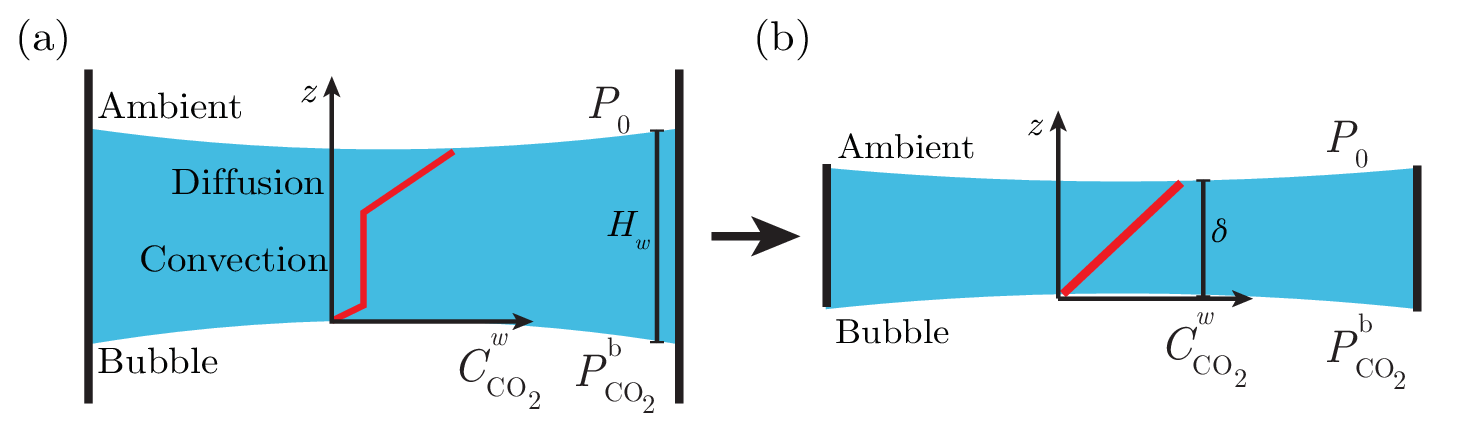}
	\caption{(a) Sketch of the quasi-steady concentration profile of CO$_2$ across the water layer after the onset of convective dissolution. The main concentration drop occurs across the top diffusion boundary layer. Transport through the convective layer is assumed instant. The concentration drop across the bottom boundary layer is neglected. This is equivalent to (b), where the water layer is treated as a single diffusion layer of thickness $\delta(t) \sim H_w \Ra(t)^{-1/4}$, where $\Ra(t)\propto [P_0-P\cb(t)]$, the  difference in the partial pressure of CO$_2$ in the CO$_2$-ambient and the bubble.}
	\label{FIG:4}
\end{figure}

At this point, we must invoke additional physical assumptions inspired by Howard's  \cite{Howard1966} and Malkus' \cite{Malkus1954} phenomenological treatment of turbulent Rayleigh--B\'enard convection.
Firstly, we assume that most of the concentration drop $\Delta C\cw$  occurs across the top diffusive boundary layer, whereas the convective region in the bulk of the water layer acts as a short-circuit for mass transfer. The concentration drop across the bottom boundary layer is assumed to be negligible. This structure is sketched in Fig.~\ref{FIG:4}(a).  Secondly, we assume that the top boundary layer must always be in a ‘marginally stable state' once it develops. This means that boundary layer cannot surpass (or shrink below) a critical or stable size $\delta(\Ra)$ which is such as to ensure that the quasi-steady mass transfer rate across the water layer is precisely given by (\ref{eq:Shmodel}). Note that the thickness of the concentration boundary layer is connected to $\Sh$ by $\delta \sim H_w/\Sh$ \cite{Grossmann2000, Bejan2013}, hence, we see that $\delta/H_w \sim  \Ra^{-1/4}$.

Under these circumstances, the quasi-steady (fully developed) concentration profile across $\delta$ can be considered linear. The quasi-steady flux  is thus $J\cwb = D\cw \Delta C\cw /\delta$. Inserting this result into the definition of $\Sh$ in (\ref{eq:RaSh}), we determine the stable quasi-stationary thickness of the boundary layer to be
  \begin{equation}
\delta(t) = \frac{H_w}{\Sh(t)} = H_w\left[ 1+ k \frac{\Ra(t)}{\Rac}\right]^{-1/4}.
\label{eq:delta}
 \end{equation}
Therefore, we ultimately treat convection as the shortening of the effective diffusion layer from $H_w$ to $\delta(t)$, as illustrated in Fig.~\ref{FIG:4}(b).
This treatment could be applied in the modelling of the various different regimes in the $\Sc$, $\Ra$ parameter space \cite{Grossmann2000} or even to situations where forced convection plays a role, in which case the governing relation in Eq. \ref{eq:Shmodel} should be adapted accordingly. In our numerical model, we simply solve the diffusion equation for $C\cw$ over the domain $0<z <\delta(t)$, where $\delta(t)$ is reevaluated at every time step.  The resulting numerical growth curves are plotted with white markers in  Fig.~\ref{FIG:3}(a) for cases (i) and (ii), with good general agreement.
Eq. (\ref{eq:delta}) also explains why the bubble growth rate still decreases with $H_w$ despite the fact that $\Ra \propto H_w^3$. Indeed, it follows that $\delta \propto H_w^{1/4}$, and a longer diffusion distance translates into weaker fluxes and slower growth.

\begin{figure}
	\centering
	\includegraphics[width=1\textwidth]{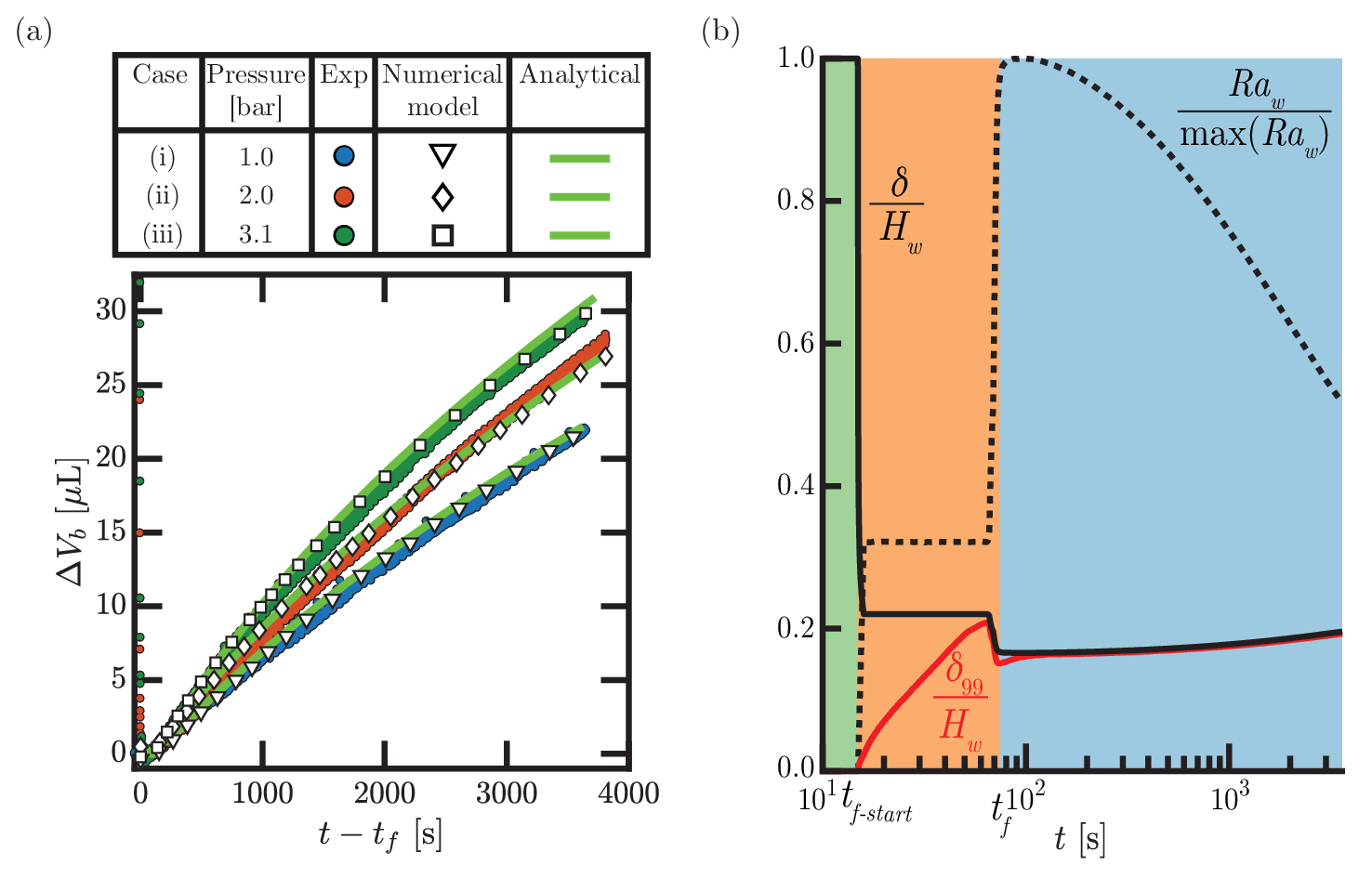}
	\caption{(a) Effect of increasing the experiment pressure level $P_0(t>t_f)$ on the bubble growth dynamics. Water layer height is $H_w \approx 4.5$ mm and post-pressurization volumes are $V_f \approx 35$~$\mu$L.  Experimental, numerical and analytical -- Eq. (\ref{eq:consol}) -- results are compared. 
	(b) Time evolution of the boundary layer thickness for marginal stability, $\delta(t)$ (black solid line), as established by its concomitant Rayleigh number (dotted line) according to the numerical model for experiment (iii) where $P_0(t>t_f) = 3.1$ bar. The red solid line represents $\delta_{99}$, the distance over which 99~\% of the numerical concentration drop takes place. Equilibrium, flushing and experiment stages are shaded accordingly  (see  Fig.~\ref{FIG:2}).}
	\label{FIG:5}
\end{figure}

The good performance of the model extends to experiments at higher pressures, as shown in Fig.~\ref{FIG:5}(a). Here, the growth rate of bubbles with post-pressurization volumes $V_f$ = 35 $\mu$L under a water layer of $H_w \approx 4.5$ mm are compared  for pressure levels $P_0(t>t_f)$ = 1.0, 2.0 and 3.1 bar. 
It becomes clear that $\dot V_b$ increases with pressure level $P_0$. This  result, perhaps counterintuitive, is purely a consequence of the stronger convection, since $\Ra \propto (P_0-P\cb) \sim P_0$. Note that in the absence of convection (pure diffusion), $\dot V_b$ remains fairly independent of $P_0$,  since both the diffusive CO$_2$ fluxes and the CO$_2$ density in the bubble are proportional to $P_0$.

The validity of the quasi-steady approximation of $\Sh(t)$ and $\delta(t)$ implied by the model in Eq. (\ref{eq:delta}) can be assessed from the time evolution of $\delta(t)$ and $\Ra(t)$  for any particular experiment. We select the  $P_0(t>t_f)$ =  3.1 bar experiment from Fig.~\ref{FIG:5}(a) for this matter; the results are shown in Fig.~\ref{FIG:5}(b). 
Before flushing and the onset of convection, $\delta/H_w = 1$. 
At the start of the flushing stage, at $t = t_\mathit{f-start}$, the critical size for stability shrinks to $\delta/H_w \approx 0.22$ in  accordance to Eq.~(\ref{eq:delta}), i.e., as established by the instantaneous $\Ra(t)$ at the flushing pressure of approximately 1 bar. Upon compression to $P_0 = 3.1$ bar shortly before $t_f$, $\delta(t)$ shrinks further accordingly. Thereafter, bubble growth and the concomitant increase in $P\cb(t)$ causes $\delta(t)$ and $\Ra(t)$  to slowly increase and decay, respectively. The slow $\delta(t>t_f) $ dynamics supports the quasi-steady approximation.

The simplicity of this approach imposes no restriction to the initial diffusive development of the boundary layer in the model. We illustrate this in Fig.~\ref{FIG:5}(b) by plotting the time evolution of $\delta_{99}$,  the distance over which 99 \% of the concentration drop takes place.
As soon as flushing begins at $t_\mathit{f-start}$, we assume that the ambient air in the tank is instantaneously replaced by CO$_2$, which sets the top boundary condition to $C\cw(H_w, \ t) = S\cw P_0(t_\mathit{f-start})$. The boundary layer evolves in a self-similar manner, and the concentration profiles can be well described by the self-similar solution 
  \begin{equation}
C\cw(z,\ t) = S\cw P_0(t_\mathit{f-start})\ \mathrm{erfc}\left(\frac{H_w-z}{\sqrt{4D\cw(t-t_\mathit{f-start})}}\right).
 \end{equation}
The diffusive boundary layer $\delta_{99}(t)$ eventually attains the critical size $\delta$, after which $\delta_{99}$ = $\delta$ and across it a linear concentration profile develops.
Pressurization at $t_f$ would of course alter the self-similar development of the profiles in the strictest sense. Nonetheless, we see that the saturation time of $\delta_{99}$ is short -- most of the $\delta_{99}$ development occurs within the short timeframe of the flushing stage itself. In fact, this saturation time, namely, the time required for the self-similar regime to transition towards the $\delta$-bounded, quasi-steady regime, can be estimated as $0.3\delta_f^2/D\cw$, where $\delta_f \equiv \delta(t_f)$ is the effective diffusion length (see appendix \ref{sec:appA}, Fig.~\ref{FIG:A1}). The concentration profiles for a previous  experiment  (where $P_0(t>t_f) = 1$ bar) drawn in  Fig.~\ref{FIG:3}(b) clearly demonstrate the rapid initial development of the boundary layer in convective experiments.
On the other hand, for diffusive experiments [see profiles in  Fig.~\ref{FIG:3}(b)], the boundary layer saturation time is $0.3H_w^2/D\cw$, i.e.,  an order of magnitude larger, which results in a noticeably longer initial transient period of slow bubble growth.

Finally, we turn our attention to the gas composition of the bubbles over time, see Fig.~\ref{FIG:3}(c).
The air (\ce{O2} and \ce{N2} gas) content in the growing bubbles appears to remain virtually unchanged throughout the duration of the experiments by virtue of the low solubility of air in water.  This observation   is especially true for the bubbles growing at  $P_0(t>t_f) = 1$ bar -- Fig.~\ref{FIG:3}(c) precisely -- since the concomitant air fluxes are weakest then. Therefore, to a good approximation, the growth in bubble volume can be fully attributed to the uptake of CO$_2$ from the water layer. This result is pivotal in the derivation of the analytical solutions for bubble growth presented in the next subsection.

\subsection{Analytical solutions for bubble growth}
\label{sec:analy}
In the analysis that follows, we consider the bubble volume dynamics for $t > t_f $, i.e., after flushing and pressurization. The initial bubble volume is thus  $V_b(t_f) \equiv V_f$. We denote the experiment pressure level $P_0(t>t_f)$ simply as $P_0$, which is treated as constant in all regards. It follows that the partial pressure of \ce{CO2} in the ambient is $P\co(t>t_f)=P_0$ precisely.
Furthermore, we assume that the flushing and pressurisation stage takes place instantaneously at $t_f$, meaning that at $t < t_f$ we consider equilibrium conditions everywhere in the system. This is reasonable since the flushing period ($t_f-t_\mathit{f-start}$) typically lasts 60 s, i.e., it is quite short in comparison to the relevant characteristic diffusion times.
 
Finally, we treat air as insoluble, i.e., we neglect the flux of \ce{O2} or \ce{N2} across the bubble or water layer. The number of moles of gas in the bubble can be then decomposed as
\begin{equation}
n_b(t) = n\cb(t) + n^b_\mathrm{N_2}(t_f) + n^b_\mathrm{O_2}(t_f),
\end{equation}
with, initially, $n\cb(t_f) = 0$.
We can then transform the above equation into a pressure--volume relationship. To do so, we make use of  the ideal gas law,  $P_0 V_b = n_b RT_0$ and  the definition of the partial pressure of CO$_2$ in the bubble, $P\cb =n\cb P_0/n_b$. Noting that $P_0 V_f = n_b(t_f)RT_0$ initially, we obtain the following relationship:
\begin{equation}
    1-  \frac{P\cb(t)}{P_0}  = \frac{V_f}{V_b(t)}.
    \label{eq:key}
\end{equation}
Next, recalling the definition of  $\Sh$ (Eq. (\ref{eq:RaSh})),  we find it useful to non-dimensionalize the net molar flux into the bubble, $\dot n_b/(\pi d^2/4)$, in the same manner. We refer to it as the dimensionless bubble growth rate:
\begin{equation}
\phi_b(t) \equiv \frac{\dot n_b(t)}{\pi d^2/4}\frac{H_w}{D\cw  \Delta C\cw (t)}.
\label{eq:phib}
\end{equation}
Substituting in the ideal gas law $\dot n_b = P_0 \dot V_b /RT_0$, using Henry's law $ \Delta C\cw (t) = S\cw [P_0- P\cb(t)]$ and relationship (\ref{eq:key}),  Eq. (\ref{eq:phib}) becomes
\begin{equation}
    \phi_b(t) = \frac{\dot V_b(t)}{Q} \frac{V_b(t)}{V_f}, \quad
    \mbox{where} \quad
    Q \equiv D\cw \frac{\pi d^2}{4 H_w}(S\cw RT_0).
    \label{eq:Shb}
\end{equation}
The dimensional quantity $Q$ can be regarded as a characteristic volume flow rate and it is a constant.
Finally, we introduce the dimensionless time coordinate, namely in multiples of the diffusion time across the water layer,
\begin{equation}
    \tau \equiv \frac{t-t_f}{H_w^2/D\cw},
    \label{eq:tau}
 \end{equation}
which will mainly serve as an argument to the transient flux functions (see  appendix \ref{sec:appA}). The transient flux functions will be shown to be fundamental in the description of the initial transient growth of the bubble. 

\subsubsection{Diffusive growth}

A fully developed (linear) concentration profile across the full height water layer $H_w$ implies, by definition, a quasi-steady $\Sh = 1$. However, there is an initial  period of boundary layer development  (lasting $\sim$ $H_w^2 /4D\cw$), during which $\Sh(t)$ evolves from 0 to 1.We can formally capture the transient rate of mass transfer  through  the inclusion of a transient correction term $f'(\tau)$ (see appendix \ref{sec:appA}). It derives from the analytical computation the diffusive flux across a layer of finite height with initial uniform zero concentration, and assuming one end of the layer (ambient interface) is at a constant concentration while the remaining end (bubble interface) remains at zero concentration. The latter assumption  that $P\cb(t)$ remains zero is completely justified given that the fractional bubble growth is indeed small during the transient period.
In such a case, 
\begin{equation}
    \Sh = 1 - f'( \tau),
    \label{eq:Shdiff}
    \end{equation}
where $f'(\tau)$ is given in  (\ref{eq:fprime}).  $\Sh(\tau)$ is plotted in Fig.~\ref{FIG:A1}(a).
Mass conservation implies that $ \phi_b = \Sh$. Equating (\ref{eq:Shb}) and (\ref{eq:Shdiff})  results in an integrable differential equation for $V_b(t)$, subject to the initial condition $V_b(t_f) = V_f$. Integration yields
 \begin{equation}
V_b(t) = V_f \sqrt{1 + \frac{2Q}{V_f}\left[ (t-t_f)- \frac{H_w^2}{D\cw} f(\tau)\right]},
\label{eq:diffsol}
 \end{equation}
 where the transient flux integral function $f(\tau)$ is defined in (\ref{eq:f}) and has limits $f(0) = 0$ and $f( \infty) = 1/6$. 
 As expected, $V_b$ does not depend on the ambient pressure $P_0$. The analytical solution (\ref{eq:diffsol}) is plotted in Fig.~\ref{FIG:3}(a) with good agreement with its experimental and numerical counterparts, (iii) and (iv).

\subsubsection{Convective growth}

The $\Sh(\Ra)$ dependence proposed in (\ref{eq:Shmodel}) can be simplified to
 $\Sh = (k \Ra/\Rac)^{1/4}$ since 
  \begin{equation}
      \Ra(t) = \frac{\lambda \cw S\cw g}{\nu_w D\cw} H_w^3\left[P_0-P\cb(t)\right]
    \label{eq:Ra_oursimp}
\end{equation}
is indeed large.
We define  $\Raf \equiv \Ra(t_f)$ as the initial (and maximum) $\Ra$ under the assumption that at $t_f$ there is still no CO$_2$ in the bubble, $P\cb(t_f) = 0$. Thereafter, $\Ra(t)$ has been shown to decay with bubble growth [cf. Fig.~\ref{FIG:5}(b)] due to the concomitant increase in $P\cb(t)$.
Combining the relationship in (\ref{eq:key}) with (\ref{eq:Ra_oursimp}) gives $\Ra(t)/\Raf = V_f/V_b(t)$. Hence, we can relate the quasi-steady Sherwood number to the bubble volume by
\begin{equation}
\label{eq:15}
 \Sh(t) =  \beta \left( \frac{V_f}{V_b(t)}\right)^{1/4}, \quad
 \mbox{with} \quad
 \beta \equiv \left( \frac{k\Raf}{\Rac}\right)^{1/4}.
  \end{equation}
The dimensionless parameter $\beta$ can be regarded as the quasi-steady Sherwood number that could be potentially attained at $t_f$ upon the immediate onset of quasi-steady convection (infinitely short transient period). Therefore, recalling that $\delta = H_w/\Sh$,  the initial critical size of the  boundary layer is set by $\delta_f = H_w/\beta$.

The initial transient regime before the onset of quasi-steady convection can be accounted for with an identical correction term as previously done in the diffusive case, except that we now consider the boundary layer development over the length $\delta_f$ instead of $H_w$. Accordingly, the argument of the transient functions $f'$ in (\ref{eq:fprime}) and  $f$ in (\ref{eq:f})  is now equal to $\beta^2 \tau$.
The transient-corrected Sherwood number becomes
 \begin{equation}
\Sh(t) = \beta \left[1 - f' ( \beta^2 \tau)\right]\left( \frac{V_f}{V_b(t)}\right)^{1/4}.
  \label{eq:Shw}
\end{equation}
Mass conservation implies $\phi_b = \Sh$. Equating  (\ref{eq:Shb}) and (\ref{eq:Shw}) followed by integration from $V_b(t_f) = V_f$ results in 
 \begin{equation}
V_b(t) =  V_f \left[ 1+ \frac{9Q}{4V_f} \beta \left\{ (t-t_f) - \frac{H_w^2/D\cw}{\beta^2}\: f(\beta^2\tau)\right\}\right]^{4/9}.
\label{eq:consol}
 \end{equation}
The analytical solution is compared with experiments and numerical solutions in Fig.~\ref{FIG:3}(a) (curves (i) and (ii)) and  Fig.~\ref{FIG:5}(a), displaying a very decent performance in all cases.
It should be pointed out that the analytical solution intentionally shares the same fitting parameter $k$ (i.e., through $\beta$) as its numerical counterpart, also bearing in mind that $k$ may vary from one experiment to another within the aforementioned bounds.

\section{Convective mass transfer: scaling}
\label{sec:scaling}

Our theoretical framework for convective dissolution is strongly based on Eq. (\ref{eq:Shmodel}) in the limit of Eq. (\ref{eq:15}) , namely, on the assumption that the quasi-steady Sherwood number depends on the instantaneous Rayleigh number as $\Sh \sim \Ra^{1/4}$.
The validity of this assumption can be verified experimentally without the need of a supporting numerical model. This requires the approximation of regarding air  as insoluble. In addition, we abbreviate the pressure level $P_0(t>t_f)$  as $P_0$, which is assumed to be constant. From earlier analyses (Sec. \ref{sec:analy}), it follows that the instantaneous Rayleigh number
\begin{equation}
    \Ra(t) = \Raf \frac{V_f}{V_b(t)}, \quad
    \mbox{where}\quad  \Raf \equiv \frac{\lambda \cw S \cw g}{\nu_w D\cw} H_w^3 P_0,
    \label{eq:Ra_final}
\end{equation}
and the dimensionless bubble growth rate defined in (\ref{eq:phib})--(\ref{eq:Shb}) (repeated here for convenience) 
\begin{equation}
    \phi_b(t) = \frac{\dot V_b(t)}{Q} \frac{V_b(t)}{V_f}, \quad
     \mbox{where}\quad  Q \equiv D\cw \frac{\pi d^2}{4 H_w}(S\cw RT_0),
    \label{eq:phib2}
\end{equation}
can both be evaluated purely from the experimental data for $t>t_f$.

\begin{figure*}
	\centering
	\includegraphics[width=1.0\textwidth]{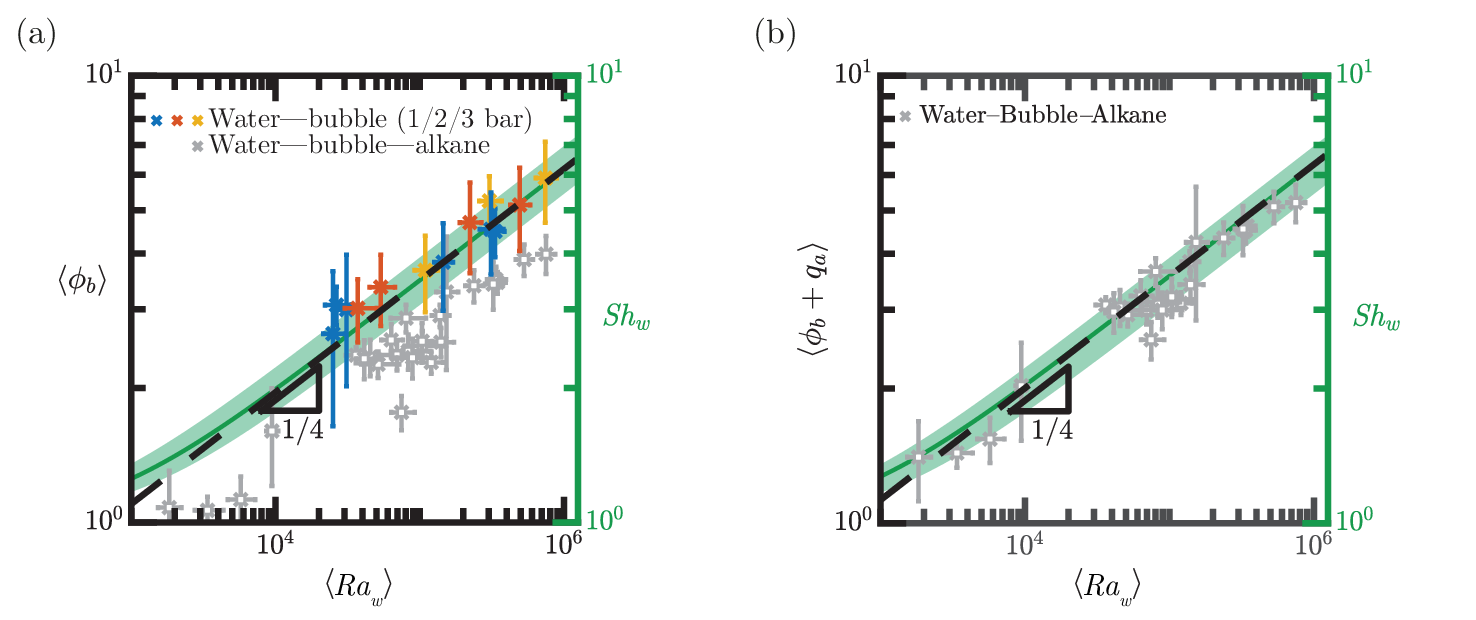}
	\caption{(a) Time-averaged dimensionless bubble growth rate as a function of the time-averaged $\Ra$  for every single water--bubble experiment (colored markers) and water--bubble--alkane experiment (grey markers). The error bars indicate the range over which $\phi_b(t)$ and $\Ra(t)$ decay during that particular experiment. The solid green line and surrounding shaded region provide the $\Sh(\Ra)$ dependence in (\ref{eq:Shmodel}): $\Sh = (1 + k \Ra/\Rac)^{1/4}$, with $k = 2.75\pm 1.25$. 
	(b) Bubble growth rate of the water--bubble--alkane experiments corrected with $q_a$, the dimensionless CO$_2$ leakage flux from the bubble to the alkane according to ({\ref{eq:qa}}). 
	}
	\label{FIG:6}
\end{figure*}

The result is shown in Fig. \ref{FIG:6}, where the time-averaged $\phi_b$ is plotted against the time-averaged $\Ra$  for every single experiment. Each colored marker represents an individual binary experiment; the grey markers belong to the ternary (water--bubble--alkane) experiments and are not expected to follow Eq. (\ref{eq:Shmodel}) as will be discussed later. The time averages are computed for $t-t_f>1100$ s to discard the initial transient growth of the bubble. The error bars indicate the range over which $\phi_b(t)$ and $\Ra(t)$ decay during that particular experiment [cf. Fig.~\ref{FIG:5}(b)].
In addition, the $\Sh(\Ra)$ dependence proposed in Eq. (\ref{eq:Shmodel}) is represented by the green solid curve, alongside its asymptotic slope of 1/4. The  shaded region around the curve delimits the range of the fitting parameter,  $k=2.75\pm1.25$, previously considered.

Figure~\ref{FIG:6}(a) shows that the binary experiments fall in the asymptotic segment of the shaded $\Sh(\Ra)$ region. This results from mass conservation, which for the binary experiments can be expressed as $\langle\phi_b\rangle = \Sh$, indicating that the CO$_2$ flux across the water layer must be equal to the CO$_2$ accumulation rate in the bubble. 
The asymptotic segment of shaded region is described by $\Sh =  a\Ra^{1/4}$, where $a = 0.2 \pm0.03$. Interestingly, our prefactor $a \approx 0.2$ seems to coincide with that obtained by \citet{Malkus1954b}  in one of the first laminar Rayleigh--B{\'e}nard  convection experiments ever conducted. It is also in complete agreement with the $0.33 \Sc^{-1/12} \approx 0.19 $ prefactor of the $I_u$ regime of the  Grossmann--Lohse theory of Rayleigh--B{\'e}nard convection in a cylindrical cell\cite{Grossmann2000}. Recent experiments of CO$_2$ bubbles growing in carbonated water \cite{Soto2019} suggest comparable values ($a \approxeq 0.3$, with the bubble radius as the relevant length scale). Dissolving sessile alcohol droplets in water \cite{Dietrich2016} yield a somewhat higher exponent ($a \approx 0.6$).
On the other hand, natural convection measurements over horizontal plates, at a quite different geometry, at $\Pra \sim 1$ result in prefactors (based on the plate diameter) approximately fourfold larger than ours \cite{Goldstein1973, Lloyd1974}.

\section{Water--bubble--Alkane system: buffering effect}

Next, we turn our attention to the fact all the ternary experiments fall systematically below the $\Sh(\Ra)$ curve in Fig.~\ref{FIG:6}(a). Indeed, $\langle \phi_b\rangle < \langle \Sh\rangle$ is a consequence of the onset of a  CO$_2$ leakage flux from the bubble into alkane layer below. 
Mass conservation now reads $\Sh = \phi_b + q_a$, where    
the (dimensionless) leakage flux $q_a(t)$ can be inferred purely from the experimental data under a set of simplifying assumptions. The result is shown in Fig. \ref{FIG:6}(b). Upon correcting $\langle\phi_b\rangle$ with $\langle \phi_b + q_a\rangle$, the experimental points are coherently shifted up into the $\Sh(\Ra)$ region. 

The expression for $q_a(t)$ is provided in Eq.~(\ref{eq:qa}) below. It will be derived in this section, which  explores the stabilising or ``buffering" effect of the alkane layer on bubble growth in greater detail.

The dissolution of \ce{CO2} into hexadecane is known to  induce density  gradients which similarly result in its enhanced transport \cite{Song2014}. 
Indeed, density measurements of hexadecane+\ce{CO2} mixtures  ($\rho_a$) \cite{Yang2019} and water+\ce{CO2}  ($\rho_w$) mixtures \cite{Mcbride2015, Garcia2001} at the same  high pressure range ($P_0>10$~MPa) and room temperature conditions coincidentally yield a solutal expansion coefficient 
\begin{equation}
\lambda\cm \equiv \left( \frac{1}{\rho_{a/w}} \frac{\partial \rho_{a/w}}{\partial C\cm}\right)_{P_0, T_0}
\end{equation} 
of approximately 10 cm$^3$/mol in both cases. Note that superscript/subscript $a$ is used to refer to the alkane (n-hexadecane) layer or medium, in the same manner as $w$ refers to water. Since $\lambda\cw$ is known to depend very weakly on $P_0$  \cite{Garcia2001} (verified by the fact that Ref. \cite{Watanabe1985} reports $\lambda\cw = 10.9$ cm$^3$/mol under atmospheric $P_0$),  it is then reasonable to assume that  $\lambda\ca\approx \lambda\cw$ also holds true in our low-$P_0$  conditions. 
 
To determine if convective dissolution across the hexadecane layer constitutes a dominant effect, we must estimate the characteristic Rayleigh number across it, $\Raa$.
At the bubble--alkane interface, the surface concentration  is $S\ca P\cb(t)$.  At the bottom of the layer a no-flux boundary condition holds. 
Assuming pure diffusive transport, and provided that the increase in $P\cb(t)$ is sufficiently slow, the maximum (or characteristic) concentration difference across the alkane layer corresponds to that at the time $t^*$ where self-similarity is broken therein, i.e. when the CO$_2$ concentration at the bottom of the alkane layer start to rise. CO$_2$ accumulates in the alkane layer thereon by virtue of the no-flux boundary condition.
The time $t^*$ is determined by $ t^*-t_f \approx 0.3 H_a^2/D\ca$ (see appendix \ref{sec:appA}). Furthermore, we approximate air as insoluble in hexadecane, whereby $P_b = P_0\Delta V_b/V_b$  from (\ref{eq:key}) applies.

The resulting expression for $Ra_a$ can be insightfully deployed in relation to the maximum Rayleigh number  across the water layer, $\Raf$ [see Eq.~(\ref{eq:Ra_final})].  Neglecting the small differences in the solubilities, diffusivities and solutal expansion coefficients of CO$_2$  in water and in hexadecane (see  \autoref{tab:constants}), we can write 
  \begin{equation}
      \frac{\Raa}{\Raf} \approx \frac{\Delta V_b}{V_b} \frac{H_a \nu_w}{H_w \nu_a}.
    \label{eq:Ra_alk}
\end{equation}
For a typical experiment with $V_f = 35$~$\mu$L and $H_a = 3$~mm, we find $ (t^*-t_f)\sim  1,200$~s and, from earlier figures, $\Delta V_b(t^*) \approx 5$~$\mu$L, hence $\Delta V_b(t^*)/V_b(t^*) \approx 1/8$. Taking $H_w \approx H_a$, and noting that $\nu_w/\nu_a \approx 1/4$, we finally obtain a characteristic magnitude of $\Raa/\Raf \sim 3$~\%. This always translates to $\Raa < 10^4$, i.e., $Ra_a \sim Ra_c$ at most. Therefore, under the current experimental conditions, diffusion can be safely assumed to remain the dominant transport mechanism across the alkane layer.
The numerical model can then be straightforwardly  extended to incorporate the presence of the alkane layer (see appendix \ref{sec:appB}), by solving the one-dimensional diffusion equation for all three species ($j=$ \ce{CO2}, \ce{O2} and \ce{N2}) over the whole height of the alkane layer. The resulting interfacial fluxes from the bubble to the alkane, $J\jba$,  modify mass conservation to $J\jwb - J\jba = 4\dot n_j^b/\pi d^2$ (cf. the discussion above Eq. (\ref{eq:Vbdot})).

\begin{figure*}
	\centering
	\includegraphics[width=0.75\textwidth]{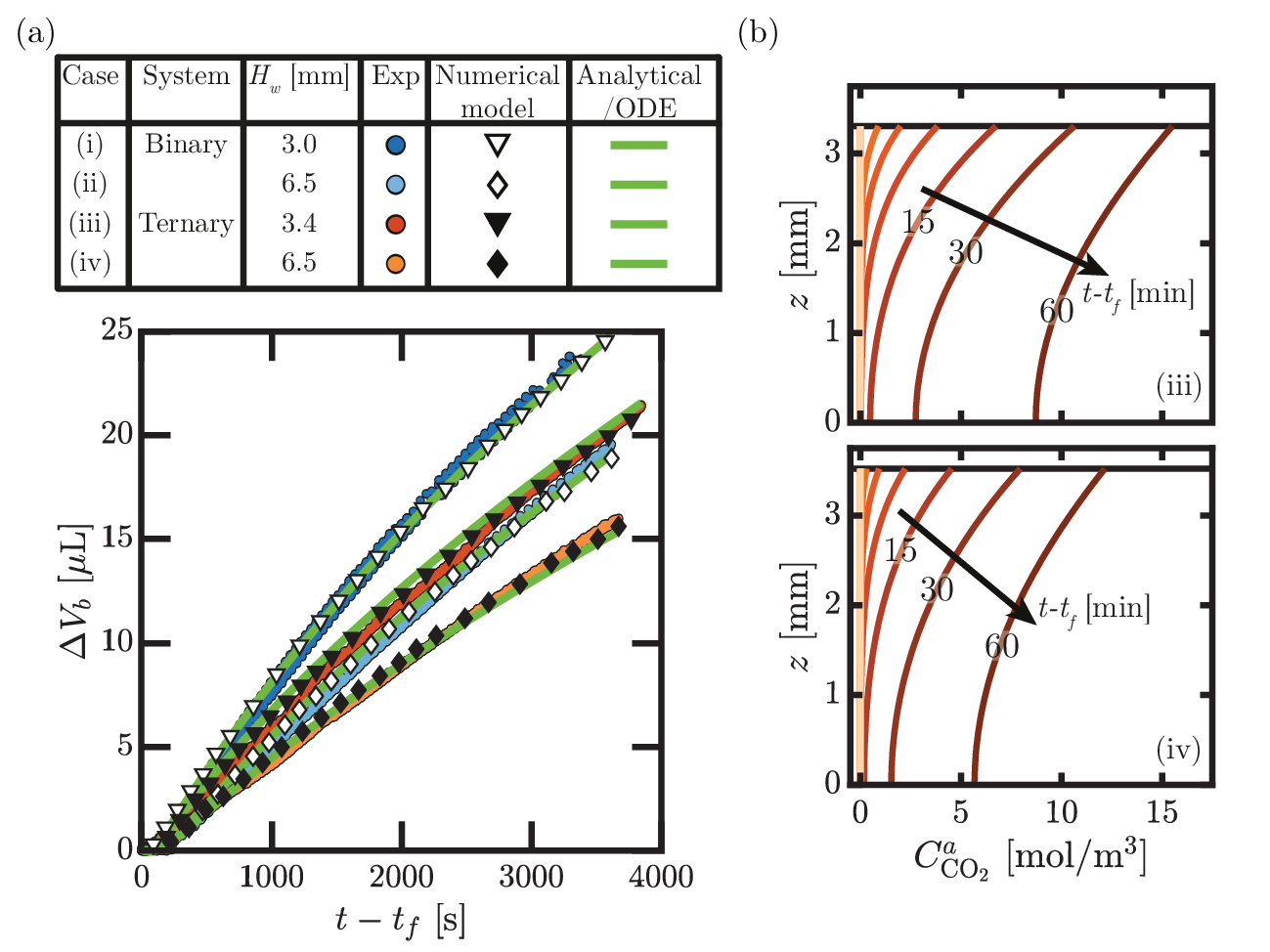}
	\caption{(a) Comparison of the bubble growth dynamics for water--bubble system (cases i, ii) and ternary water--bubble--alkane systems (cases ii, iv) for two distinct water layer heights.  The alkane layer height is $H_a \approx 3.5$ mm. In all four cases, $V_f$ = 35 $\mu$L and $P_0 = 1$ bar. Experiments (i, ii) are overlaid with numerical and analytical (\ref{eq:consol}) solutions;  experiments  (iii, iv) with numerical and ODE (\ref{eq:ODE}) solutions.  
	(b) Time evolution of the CO$_2$ concentration profile in the alkane layer according  to the numerical model for case (iii) [top panel] and case (iv)[bottom panel].}
	\label{FIG:7}
\end{figure*}

Figure \ref{FIG:7}(a) compares the bubble growth dynamics of two binary experiments at $P_0(t>t_f) = 1$ bar [same cases (i) and (ii)  from Fig.~\ref{FIG:3}] with their equivalent ternary systems with $H_a \approx 3.5$ mm under otherwise the same conditions.  
The buffering effect of the n-hexadecane layer on bubble growth first introduced in Fig.~\ref{FIG:2} is clearly observed again.
The good performance of the numerical solution validates, once more, the assumption of negligible convective dissolution across the alkane layer. On another note, the transient nature of $J\jba(t)$ can be inferred from the time evolution of the numerical CO$_2$ concentration profiles in the alkane layer shown in Fig.~\ref{FIG:7}(b). 
 Quasi-steady or self-similar profiles or fluxes are never attained. This is a consequence of  the unsteady interfacial concentration $C\ca(H_a, \ t)$   being strictly an increasing function of time and the no-flux boundary condition on $z = 0$. Notice that CO$_2$ starts accumulating in the alkane layer after a time of approximately 15 minutes, in accordance with our previous estimation.
 
The added analytical complexity brought by the transient nature of the CO$_2$ transport across the alkane layer can be witnessed, for example, in the unmanageable time-integral-laden mathematical expression of the concentration profile across a finite sheet subject to a no-flux boundary condition and variable surface concentration \cite{Carslaw1959}.
However, noting that on the time scale of our experiments the hexadecane layer  hardly becomes saturated [cf. Fig.~\ref{FIG:7}(b)], we may choose to ignore the effect of the no-flux boundary. We do so by treating the hexadecane layer as a semi-infinite medium. Not only is this a reasonable simplification, it also poses a problem of particular interest. In such a case, and treating air as insoluble, we can reduce the model formulation to a single ordinary differential equation for $V_b(t)$ as will be shown in the next subsection.

\label{sec:wba}
\subsection{Dynamic equation for bubble growth}
 Approximating air as insoluble renders Eq.~(\ref{eq:key}) which allows us to relate the surface concentration of CO$_2$ in the alkane layer at the alkane--bubble interface to the instantaneous bubble volume by
\begin{equation}
C_s(t) = S\ca P\cb(t) = {S\ca P_0} \left[ 1-\frac{V_f}{V_b(t)}\right],
\label{eq:cs}
\end{equation}
where $P_0$ here abbreviates the constant pressure level $P_0(t>t_f$).
It should be noted that, unlike in water, the solubilities of air and CO$_2$ in n-hexadecane are quite comparable (see \autoref{tab:constants}). 
Therefore, the approximation of neglecting the flux of air into the hexadecane layer is mostly valid for $P_0 = 1$ bar because the hexadecane layer is initially saturated with air at that pressure.  Pressurisation beyond this value results in the immediate undersaturation of the later;
raising $P_0$ further eventually yields fluxes of air and CO$_2$  of similar magnitude. 

Next, treating the alkane layer as semi-infinite fully removes the length scale $H_a$ from the problem, which allows us to conveniently work with the water-layer time $\tau$ defined in Eq.~(\ref{eq:tau}) instead.
The CO$_2$ diffusion flux from the bubble to the semi-infinite alkane layer with a time-dependent surface concentration is given by (see e.g. Ref. \cite{Landau1987})
\begin{equation}
J\cba = \frac{D\ca}{H_w} 
\left[\frac{1}{\sqrt{\pi}} \int_0^\tau\frac{1}{\sqrt{\tau-u}} 
\frac{\dd C_s}{\dd u} \:\dd u \right]
\label{eq:history}
\end{equation}
Identical history integral flux terms to the one above can be found deployed in previous analytical treatments concerning the  growth and dissolution of drops/bubbles with unsteady surface concentrations \cite{Penas2016, Chu2016}. The integral describes the so-called history effect, i.e., the contribution of past mass transfer events on the current  growth dynamics.
The fact that the recent history contributes the most,  $1/\sqrt{\tau-u} \rightarrow \infty$ as $u \rightarrow \tau$, can be used to evaluate the time integral in our case. Since the bubble growth dynamics are quite steady (no abrupt or fast changes in $\dot V_b$), we can approximate ${\dd C_s}/{\dd u}$ as  the current time derivative $\dd C_s(\tau)/\dd \tau$ irrespective of $u$.  Analytical integration is then possible, and the flux becomes
\begin{equation}
J\cba = \frac{D\ca}{H_w} 
\left[2\sqrt{\frac{\tau}{\pi}}\: \frac{\dd C_s}{\dd \tau} \right]
= \left(\frac{2 D\ca S\ca P_0 V_f}{\sqrt{\pi D\cw}}\right) \frac{\dot V_b(t)}{V_b(t)^2} \sqrt{t-t_f},
\label{eq:Jba}
\end{equation}
where use of Eq.~(\ref{eq:tau}) and Eq.~(\ref{eq:cs}) has been made.
The dimensionless flux, $q_a$, into the alkane (normalised by the water-layer diffusive flux,  in coherence with $\Sh$ and $\phi_b$) is 
\begin{equation}
q_a(t) \equiv \frac{J\cba(t) H_w}{D\cw  \Delta C\cw (t)}
=  \sqrt{T^{aw}}\sqrt{t-t_f}\  \frac{\dot V_b}{V_b}
, \quad
\mbox{with} \quad
 \sqrt{T^{aw}} =\frac{2 H_w}{ \sqrt{\pi D\cw}}\frac{D\ca}{D\cw} \frac{S\ca}{S\cw}.
\label{eq:qa}
\end{equation}
Here $T^{aw}$ is a time constant which depends purely on the system properties.  The remaining ingredients have been derived in Sec. \ref{sec:analy}; these are naturally $\phi_b$  and $\Sh$, whose expressions are given in Eq.~(\ref{eq:phib}) and  Eq.~(\ref{eq:Shw}) respectively.
Finally, mass conservation  yields   
\begin{equation}
\phi_b = \Sh - q_a.
\end{equation}
Substituting the expressions in Eq.~(\ref{eq:phib}), Eq.~(\ref{eq:Shw}) and Eq.~(\ref{eq:qa}),  the mass balance becomes
\begin{equation}
 \frac{\dot V_b}{Q} \frac{V_b}{V_f} =
 \left\{  \beta \left[1- f'(\beta^2\tau)\right]\left( \frac{V_f}{V_b}\right)^{1/4} \right\}
  - \left\{ \sqrt{T^{aw}}\sqrt{t-t_f}\  \frac{\dot V_b}{V_b}\right\},
  \label{eq:ODE}
 \end{equation}
which constitutes an ordinary nonlinear differential equation for $V_b(t>t_f)$
that can be numerically integrated subject to the initial condition $V_b(t_f) = V_f$.
The solution to (\ref{eq:ODE}) is plotted in Fig.~\ref{FIG:7}(a) for cases (iii) and (iv), where the ODE solution shares the same fitting coefficient $k$ as its numerical counterpart. Overall, excellent agreement is observed, despite the underlying simplifications. We also point out that the flux $q_a(t)$ in Eq.~(\ref{eq:qa}) is independent of $k$ (or any other fitting parameter) and it can be evaluated directly from the experimental measurements, as done for Fig. \ref{FIG:6}(b).

\begin{figure*}
	\centering
	\includegraphics[width=0.85\textwidth]{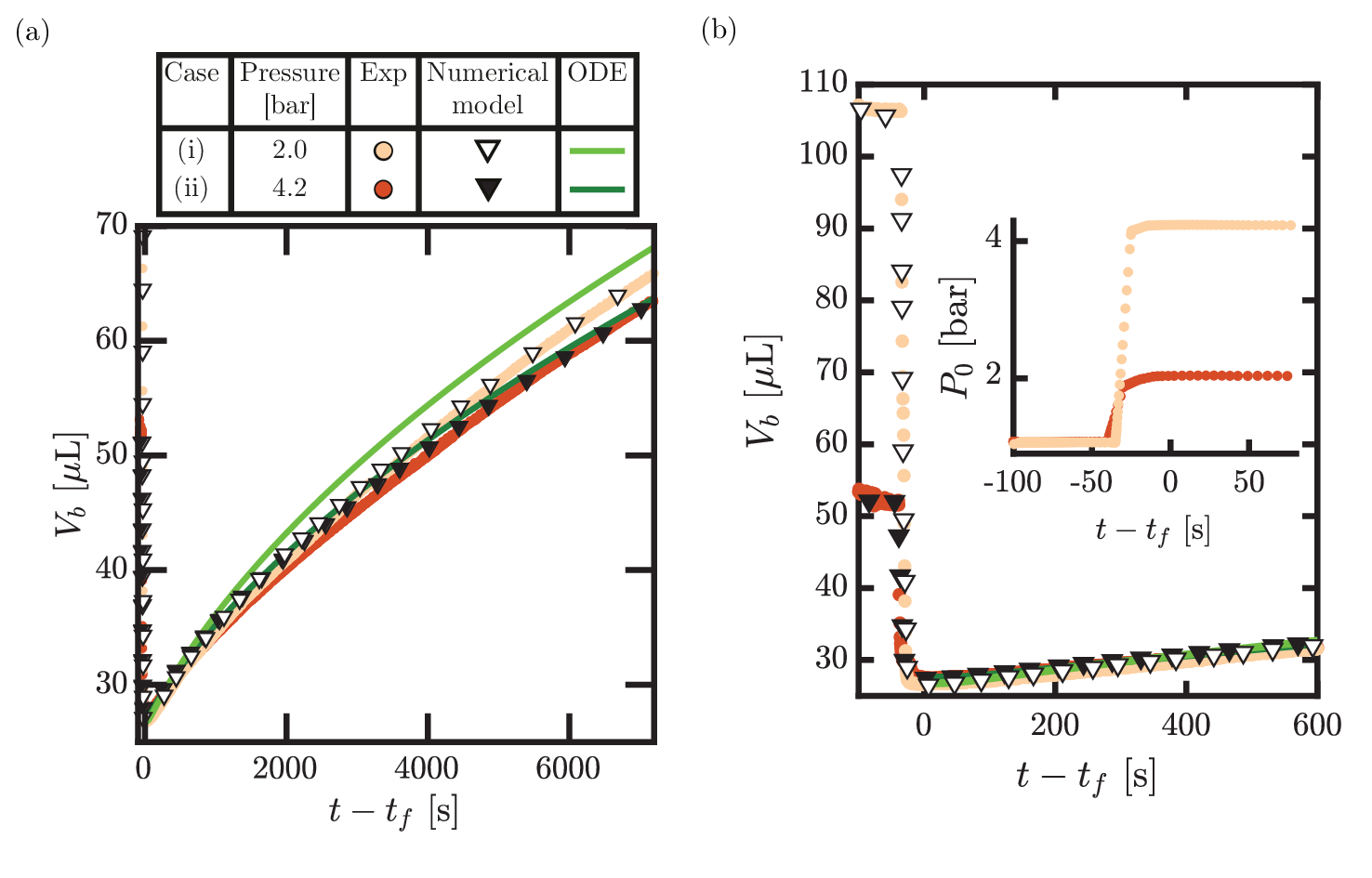}
	\caption{(a) Bubble growth dynamics for the water--bubble--alkane system at higher pressure levels ($V_f = 27 \mu$ L). Case (i): $P_0= 2.0$ bar, $H_w = 3.0$ mm, $H_a = 3.2 $ mm; 
	 case (ii)  $P_0= 4.2$ bar, $H_w = 3.0$ mm, $H_a = 3.2 $ mm. Experimental, numerical and ODE (\ref{eq:ODE}) results are compared.
(b) Close-up of the pressurisation stage and initial growth. Inset: Concomitant pressure time history.}
	\label{FIG:8}
\end{figure*}

 Figure~\ref{FIG:8}(a) shows the bubble growth dynamics of the water--bubble--alkane system ($H_w \approx H_a \approx 3 $ mm) at higher experiment pressure levels, namely at $P_0= 2.0$ and 4.2 bar.
Interestingly, the alkane buffering effect also visibly diminishes the dependence of the growth rate $V_b$ on $P_0$.  The separation of the growth curves is indeed less pronounced than in the binary system, see Fig. \ref{FIG:5}(a).
The limitation of the ODE solution becomes apparent at these higher pressures. The ODE  overestimates the growth rate, a consequence of neglecting the dissolution flux of air from the bubble into the alkane. The deviation worsens with increasing $P_0$, i.e., with the air flux magnitude.
On another note, the accuracy of the numerical model in reproducing the initial pressure-bound dynamics is highlighted in  Fig.~\ref{FIG:8}(b), which  shows a close-up of the initial minutes of bubble growth and the preceding pressurisation stage. Bubble dissolution for $t>t_f$ is not observed, despite the fact that the dissolution flux of air into the alkane is strongest at $t \approx t_f$. The onset of convective dissolution is as fast as the time required for the step in pressure to reach and stabilise at the experiment level [see Fig.~\ref{FIG:8}(b) inset].

\subsection{Nitrogen solute exchange}

\begin{figure}
	\centering
	\includegraphics[width=0.48\textwidth]{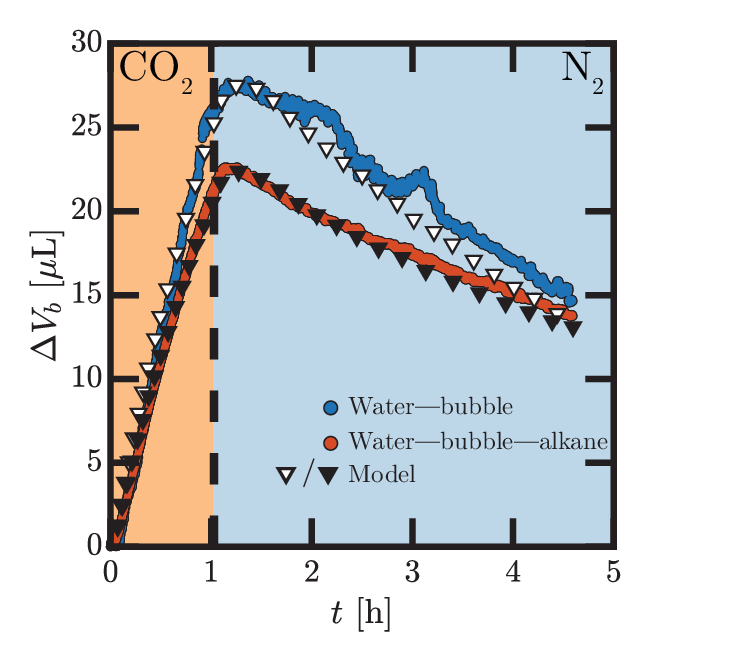}
	\caption{Bubble growth dynamics for a binary and ternary system ($V_f$ = 35 $\mu$L, $H_w$ = 3 mm and $H_a$ = 3 mm) at $P_0 =1$ bar. After $t = 1$ h, the CO$_2$ atmosphere inside the chamber is replaced by a N$_2$ atmosphere at the same pressure, thereby enforcing dissolution. }
	\label{FIG:9}
\end{figure}

To further visualise the buffering effect of the hexadecane layer on bubble dissolution, a two-cylinder experiment consisting of a binary and ternary system ($V_f$ = 35 $\mu$L, $H_w$ = 3 mm and $H_a$ = 3 mm) was placed in a CO$_2$ atmosphere at $P_0(t>t_f)$ = 1.0 bar.  After 1 hour, the CO$_2$ atmosphere is replaced by a  N$_2$ atmosphere at the same pressure. The resulting bubble dynamics are shown in  Fig.~\ref{FIG:9}.
The first solute exchange event induces bubble growth, as discussed extensively in the previous sections. In contrast, the second solute exchange induces bubble dissolution,  as CO$_2$ from the bubbles redissolves back through the water barrier. Note that the opposing  N$_2$ flux is much weaker owing to its poor solubility in water. The buffering effect of the alkane layer stands out with the fact that the bubble in the ternary system shrinks slower compared to the bubble in the bubble--water system. This difference is caused by CO$_2$ exsolving from the alkane back into the bubble. 
After replacing the CO$_2$ atmosphere with N$_2$, the CO$_2$ concentration gradient across the water layer is effectively inverted. The dense CO$_2$-rich boundary layer now rests at the bottom water--bubble interface. Stratification is stable thereon, and therefore the transport of CO$_2$ can be assumed to be purely driven by diffusion. This evidently results in a slower dissolution rate. The model prediction importantly reflects the asymmetry between the growth and dissolution stages, and overlaps reasonably well with the experimental results.

\section{Conclusions \& Outlook}
\label{sec:conc}

We have investigated the growth dynamics of a trapped slug bubble in a vertical glass cylinder beneath a short layer of water which separates the bubble from the ambient gas. Replacing the ambient air with a CO$_2$ atmosphere induces an asymmetric exchange of the gaseous solutes between the CO$_2$-rich water barrier and the air-rich bubble.  Net bubble growth is always observed, even after pressurisation. We refer to this process as solute exchange. 

The dominant transport of  CO$_2$ across the water barrier is facilitated by diffusion and greatly enhanced by convective dissolution.  
Inverting the cylinder orientation has correspondingly been shown to suppress convection. The transport of CO$_2$ across the layer was then fully driven by diffusion, resulting in slower growth.
The bubble growth dynamics and underlying mass transport processes were characterised by means of a numerical model based on one-dimensional diffusive transport. Analytical solutions able to accurately predict the growth dynamics were subsequently derived. 
The effect of convective dissolution across the water layer was treated as a reduction of the effective diffusion length, in accordance with the expected scaling for laminar or natural convection. The fact that the Sherwood number scales to the 1/4 power of the Rayleigh number was verified experimentally, and explains why the bubble growth rate was observed to increase with the ambient pressure but behaves inversely with the  the water layer height.

Finally, the binary water--bubble system was compared to a ternary water--bubble--alkane system.
The alkane (n-hexadecane) layer was shown to act essentially as an unsteady sink (or source) of CO$_2$ gas. As a result, the alkane layer induces a buffering effect on the bubble growth (or dissolution) dynamics. The slower growth curves were well reproduced by an extension of the numerical model and ultimately a single dynamical equation for the bubble volume derived under a set of simplifying assumptions.

Our findings offer new insight on mass-transfer effects in microfluidic or microreactor devices comprising segmented gas--liquid phases or  density-changing solutes, and also on the growth and elimination of trapped bubbles. Moreover, the differences in the growth rates between binary systems, ternary systems, or a combination thereof, can offer a feasible means by which the physical properties of gases in the liquid layers can be obtained.

\section*{Acknowledgements}
This work was supported by the Netherlands Center for Multiscale Catalytic Energy Conversion (MCEC), an NWO Gravitation programme funded by the Ministry of Education, Culture and Science of the government of the Netherlands.

\appendix

\section{Diffusion across a sheet of finite length}
\label{sec:appA}
Consider the canonical problem of one-dimensional diffusion across a sheet of finite length. One end of the sheet is kept at a constant surface concentration while the remaining end is at zero concentration. The sheet concentration profile is initially zero. 
With proper normalisation, the dimensionless problem considers a surface concentration,  diffusivity and sheet length of unity. Our particular case involves the diffusion of  $C\cw(z, \ t>t_f)$ across a water layer of height $H_w$ with surface concentrations $C\cw(H_a, \ t) = S\cw P_0$ and  $C\cw(0, \ t) =0$. We then consider $c=C\cw/S\cw P_0$ as a function of $\xi = 1-z/H_w$ and $\tau = D\cw(t-t_f)/H_w^2$.

\begin{figure}
	\centering
	\includegraphics[width=0.7\textwidth]{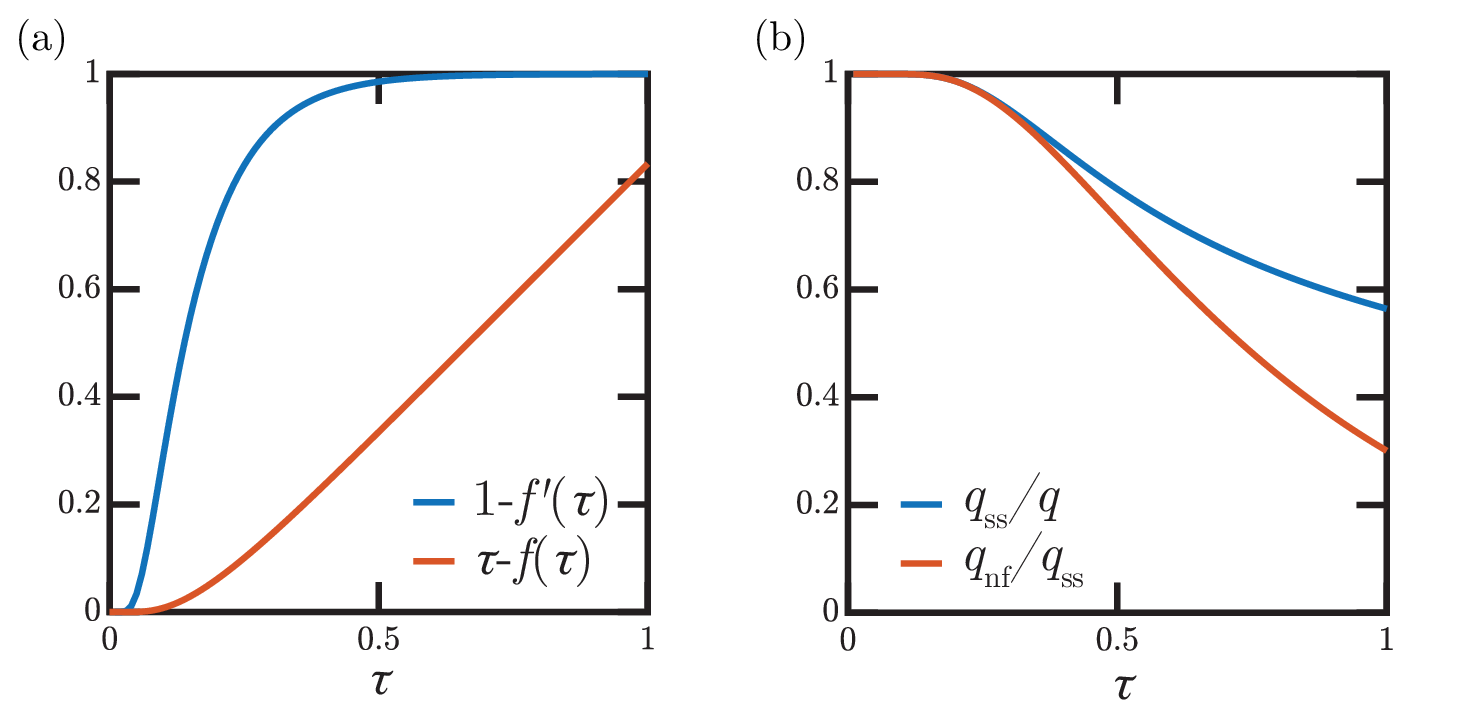}
	\caption{(a) Diffusion outflux  (blue line) and its time integral (red line) as functions of time. The transient period can be thought to last $\tau \sim0.4$, after which the flux overcomes the 95~\% mark of the steady-state value.
	(b) Influx ratios:  $q_{ss}$ denotes the (self-similar) diffusion influx into the unit-concentration boundary of an infinite sheet,  $q$  is the  influx into the unit-concentration boundary of a sheet of unit length delimited by a zero-concentration boundary, and $q_\mathit{nf}$ is the influx into the unit-concentration boundary of a sheet of unit length delimited by a no-flux boundary. In both cases, self-similarity breaks down after a time  $\tau \sim0.3$, where the flux ratios drop below 95~\%. }
	\label{FIG:A1}
\end{figure}

The solution to $ \partial_\tau c =  \partial_{\xi\xi} c$ subject to $c(0, \ \tau) = 1$, $c(1, \ \tau) = 0$ and $c(\xi, \ 0) = 0$ is (see e.g. Ref. \cite{Crank1975})
\begin{equation}
c(\xi, \ \tau) =  (1-\xi)
- \frac{2}{\pi} \sum_{n=1}^{\infty} \frac{1}{n} \sin (n \pi \xi)
\ \ee{-n^2 \pi^2 \tau }.
\label{eq:cseries}
\end{equation}
Fluxes can be easily deduced from (\ref{eq:cseries}).
At any given time, the diffusion flux out of the sheet (across the zero-concentration end) is  $-\left.\partial_\xi c(\xi, \ \tau)\right|_{\xi=1} = 1-f'(\tau)$, where
 \begin{equation}
f'(\tau) = -2 \sum_{n=1}^{\infty} (-1)^n \ee{-n^2 \pi^2 \tau}
\label{eq:fprime}
\end{equation}
is referred to as the transient flux function. 
It has limits $f'(\infty) = 0$ and $f'(\tau\rightarrow 0) =1$ but $f'(0)$ is non-analytical in $\tau = 0$.
The diffusion flux $1-f'(\tau)$ is plotted in  Fig.~\ref{FIG:A1}(a) and it is equivalent to our Sherwood number given that it precisely quantifies mass transfer out of the sheet. This result is directly used in (\ref{eq:Shdiff}).

Figure \ref{FIG:A1}(a) also plots the total amount of substance that has diffused out in time $\tau$, namely $\tau- f(\tau)$, where
 \begin{equation}
f(\tau) \equiv \frac{1}{6} +\frac{2}{\pi^2} \sum_{n=1}^{\infty} \frac{(-1)^n}{n^2} \ee{-n^2 \pi^2 \tau }, 
\label{eq:f}
\end{equation}
is referred to as the transient flux integral function. It has limits  $f(0) = 0$, and  $f(\infty) = 1/6$.

In the limit  $\tau \rightarrow 0$,  (\ref{eq:cseries}) converges to the self-similar solution 
$c_{ss}(\xi, \ \tau) =  \erfc (\xi/ \sqrt{4\tau})$.
The time at which self-similarity is broken can be estimated as the time at which the finite length of the sheet causes the influx (on $\xi = 0$) to deviate from the self-similar solution by an arbitrary fractional amount
This is done in  Fig.~\ref{FIG:A1}(b), which plots the ratio of the influx of the self-similar solution, $q_{ss} = 1/\sqrt{\pi \tau}$, to the influx of the solution in (\ref{eq:cseries}), namely
\begin{equation}
q =  1+2\sum_{n=1}^{\infty}
\ \ee{-n^2 \pi^2 \tau }.
\end{equation}
This result is relevant  in the description of the diffusive transport in the water layer. In the case of the alkane layer, we should consider the diffusion across an equivalent sheet where the zero-concentration boundary  $c(1, \ \tau) = 0$ is replaced by a no-flux boundary, i.e.,  $\partial c/\partial \xi = 0$ on $\xi = 1$. The influx (on $\xi = 0$) becomes  (see e.g. Ref. \cite{Crank1975})
\begin{equation}
q_\mathit{nf} = 2 \sum_{n=0}^\infty
  \exp\left[-\pi^2\left(n+\frac{1}{2}\right)^2 \tau\right] .
  \label{eq:cgrad2}
  \end{equation}
 Similarly, the time at which self-similarity is broken can be estimated as the time in which $q_\mathit{nf}/q_{ss}\leq 1$ falls below an arbitrary threshold, see  Fig.~\ref{FIG:A1}(b).

\section{One-dimensional numerical model}
\label{sec:appB}

Following the notation of main text, the quantities associated with the water phase are indicated by subscript/superscript $w$; subscript/superscript $b$  refers to the bubble, $a$ to the alkane and $0$ to the ambient.
 The transport of molar concentration of gas species $j =$ CO$_2$, O$_2$ or N$_2$ dissolved in the water layer is modelled by the 1D diffusion equation,
\begin{equation} \label{eq:diffeq}
	\frac{\partial \cwj}{\partial t} 
	= \Dwj  \frac{\partial^2 \cwj}{\partial x^2},
	\quad \mbox{on}\quad  0<x<H_w .
\end{equation}
The concentration boundary conditions in the water phase follow Henry's law,
\begin{equation}
\label{eq:bc}
	\cwj(0, \ t) = \kwj \pbj(t),
	\quad
	\cwj(H_w, \ t) = \kwj X_j^0(t) P_0(t) ,
	\end{equation}
where the ambient pressure $P_0(t)$ and ambient mole fractions, $X_j^0(t)$ are prescribed at all times. In our case, the pressure history $P_0(t)$ is directly interpolated from the experimental measurements.
In the presence of dissolution-driven convection, we must solve for $C\cw$ within a single diffusive region (boundary layer) of length $\delta(t)$ as explained in the main text, that is
\begin{equation}
	\frac{\partial C\cw}{\partial t} 
	= D\cw \frac{\partial^2 C\cw}{\partial \xi^2}
	\quad \mbox{on}\quad  0<\xi<\delta(t) .
	\label{eq:diffco2}
\end{equation}
The numerical difficulties inherent in solving the moving-boundary problem posed in  (\ref{eq:diffco2})  can be circumvented by introducing a coordinate transformation $x(t) = H_w \xi/2\delta(t)$ which reverts the spatial domain back to $0 \leq x \leq H_w$. We then solve for $C\cw(x, \ t)$ from
\begin{eqnarray}
	\frac{\partial C\cw}{\partial t} - \frac{\dot \delta}{\delta}x \frac{\partial C\cw}{\partial x}
	= D\cw \left(\frac{H_w}{\delta}\right)^2
	\frac{\partial^2 C\cw}{\partial x^2} \quad
\mbox{on}\quad  0<x <H_w .
\label{eq:numdiffeq}
\end{eqnarray}
The boundary conditions in (\ref{eq:bc}) therefore still apply.
Note that a numerical advection term [second term in the left-hand-side of (\ref{eq:numdiffeq})] formally appears as a consequence of the coordinate $x(t)$ changing with time. It is however neglected under the assumption that the variations in $\delta$ are slow as compared to diffusion, i.e., $ \dot \delta \delta/D\cw\ll 1$. 
The thickness $\delta$ is computed at every time step  as established by the relationship in (\ref{eq:delta}), where the numerical value of $\Ra(t)$ [as defined in (\ref{eq:RaSh})] is evaluated at the previous time step.

When modelling the ternary water--bubble--alkane system,  we must additionally solve the diffusion equation in the alkane layer,
\begin{equation}
	\frac{\partial \caj}{\partial t} 
	= \Daj\frac{\partial^2 \caj}{\partial y^2},
	\quad \mbox{on}\quad  0<y< H_a,
\end{equation}
with boundary conditions
\begin{equation}
	\frac{\partial \caj}{\partial y}(0, \ t) = 0,
	\quad
	\caj(H_a, \ t) = \kaj \pbj(t).
\end{equation}
The partial pressures inside the bubble are assumed proportional to the mole fractions, $\pbj = \nbj P_b/n_b$, where  $n_b(t) = \sum_j \nbj(t)$ is the total number of moles of gas and the assumption  $P_b(t) = P_0(t)$  neglects the Laplace pressure in the bubble. The gas content in the bubble is found by integrating  in time the molar fluxes into the bubble,
\begin{equation}
	\frac{4}{\pi d^2}\frac{\dd \nbj }{\dd t} = 
	\Dwj\frac{\partial \cwj}{\partial x}(x=0, \ t)-\Daj\frac{\partial \caj}{\partial y}(y=H_a, \ t). 
\end{equation}
Finally, the bubble volume is assumed to evolve according to the ideal gas law,
$P_0(t)V_b(t) =  n_b(t) R T_0$.
Before flushing ($t < t_\mathit{f-start}$), the system is in thermodynamic equilibrium. The initial bubble volume is $V_b(t<t_\mathit{f-start}) = \Veq$ and $ P_0(t<t_\mathit{f-start})=\peq$. The initial conditions read 

\begin{eqnarray}
	\caj(x, \ t<t_\mathit{f-start}) = \kaj \peq X_j\oeq,\nonumber\\
\cwj(y, \ t<t_\mathit{f-start}) = \kwj X_j\oeq\peq 
\end{eqnarray}
and
\begin{equation}
	\nbj(t<t_\mathit{f-start}) = \frac{\peq X_j\oeq\Veq}{RT_0}
\end{equation}
where
$X_{\ce{CO2}}\oeq= 0$,  
$X_{\ce{N2}}\oeq = 0.79$ and
$X_{\ce{O2}}\oeq = 0.21$ is the composition of dry air.
Immediately upon flushing ($t > t_\mathit{f-start}$), the ambient composition is set to reflect a pure CO$_2$ atmosphere: $X_{\ce{CO2}}^0 = 1$, and $X_{\ce{N2}}^0 = X_{\ce{O2}}^0 = 0$. 
The diffusion equations are discretized using a second-order central finite-differences scheme. The resulting ODE system for  $\cwj(x,\ t)$, $\caj(y,\ t)$ and $\nbj (t)$ is integrated in time with a standard Runge--Kutta ODE solver.

\bibliography{BGSE_PRFV2}

\end{document}